\documentclass[12pt]{iopart}
\usepackage{iopams}

\expandafter\let\csname equation*\endcsname\relax

\expandafter\let\csname endequation*\endcsname\relax

\usepackage{amsmath}
\usepackage{hyperref}

\usepackage[english]{babel}
\usepackage{graphicx}
\usepackage[dvipsnames,table]{xcolor}

\newcommand{\op}[1]{{\sf #1}}
\newcommand{\ox}{\op{x}}
\newcommand{\oc}{\op{c}}
\newcommand{\oP}{\op{P}}
\newcommand{\oM}{\op{M}}
\newcommand{\oH}{\op{H}}
\newcommand{\oU}{\op{U}}

\newcommand{\oO}{\op{O}}
\newcommand{\opsi}{\hat{\psi}}
\newcommand{\ochi}{\hat{\chi}}
\newcommand{\cL}{{\cal L}}
\newcommand{\vect}[1]{\textbf{\textit{#1}}}
\newcommand{\la}{\langle}
\newcommand{\ra}{\rangle}

\newcommand{\da}{{\dagger}}
\newcommand{\diff}{{\rm d}}
\newcommand{\qerw}[1]{\left\langle #1 \right\rangle}
\newcommand{\qerwsm}[1]{\langle #1 \rangle}

\begin{document}

\title[Sensing spontaneous collapse and decoherence with interfering BECs]{Sensing spontaneous collapse and decoherence with interfering Bose-Einstein condensates}
\author{Bj\"{o}rn Schrinski$^1$, Klaus Hornberger$^1$, Stefan Nimmrichter$^2$}
\address{$^1$ Faculty of Physics, University of Duisburg-Essen, Lotharstraße 1-21, 47048 Duisburg, Germany}
\address{$^2$ Centre for Quantum Technologies, National University of Singapore, 3 Science Drive 2, Singapore 117543}

\begin{abstract}
We study how matter-wave interferometry with Bose-Einstein condensates is affected by hypothetical collapse models and by environmental decoherence processes. Motivated by recent atom fountain experiments with macroscopic arm separations, we focus on the observable signatures 
of first-order and higher-order coherence for different two-mode superposition states, and on their scaling with particle number.
This can be used not only to assess the impact of environmental decoherence on many-body coherence, but also to quantify the extent to which macrorealistic collapse models are ruled out by such experiments.
We find that interference fringes of  phase-coherently split condensates are most strongly affected by decoherence, whereas the quantum signatures of independent interfering condensates are more immune against macrorealistic collapse. A many-body enhanced decoherence effect beyond the level of a single atom can be probed if higher-order correlations are resolved in the interferogram.

\end{abstract}

\maketitle


\section{Introduction}
Recent years have witnessed the experimental demonstration of quantum superposition states far beyond the atomic regime.
In particular, experiments delocalizing mechanical degrees of freedom, vie for the sheer mass or number of constituent particles involved.
Center-of-mass interference with composite nano-objects  is well established \cite{Arndt1999,
Gerlich2011,hornberger2012colloquium}, and experimental efforts to  cool the  vibrational modes of micromechanical resonators or levitated particles are approaching a regime where quantum coherence  may become observable
\cite{teufel2011sideband,chan2011laser,Aspelmeyer2014,peterson2016laser}.
Collective quantum coherence can also be observed in interference experiments with dilute and weakly interacting many-body systems, such as Cooper-paired electrons in superconductors \cite{Friedman2000,vanderWal2000,Hime2006} or Bose-condensed atoms coherently split between two spatial modes \cite{Shin2004,Jo2007,Baumgaertner2010,Schmiedmayer2013,Schmiedmayer2014}. Here, number squeezing can lead to a significant amount of many-body entanglement between the two modes of the condensate 
\cite{Jo2007,Schmiedmayer2013}.
At the same time, single-atom interference experiments have reached unprecedented degrees of precision and enormous scales in terms of interference arm lengths \cite{Muller2008,Chiow2011,McDonald2013} to the extent that even tiny gravitational forces and relativistic corrections of the Schr\"{o}dinger equation can be detected \cite{Peters1999,Borde2002,Dimopoulos2007}.

These experiments all have in common that they push the domain of quantum mechanics far into the macroscopic regime with respect to one or another figure of merit---be it mass, arm separation, or coherence time \cite{Hornberger2014}. They are thus raising the stakes for proponents of objective collapse theories and macrorealism \cite{Leggett2002,Bassi2013}. In fact, this gives rise to an objective, empirical method to quantify the degree of macroscopicity attained in 
mechanical quantum superposition experiments, since one experiment can be deemed more macroscopic than another if it rules out a greater set of macrorealistic modifications to quantum theory. This can be turned into a quantitative statement by specifying the mathematical form of a generic class of such modifications, as derived from a number of basic symmetry and consistency requirements \cite{Nimmrichter2013}. We will refer to them as \emph{minimal macrorealistic modifications} (MMM) in the following. The model of Continuous Spontaneous Localization \cite{Ghirardi1990,Bassi2003,Bassi2013} is a renowned special case, whose impact on optomechanics and matter-wave experiments is being scrutinized \cite{nimmrichter2011testing,Bahrami2014,Nimmrichter2014,Diosi2015,Vinante2016,Bilardello2016,Bilardello2017}.
In turns out that, by this empirical standard of macroscopicity, the latest atom interferometers are on par with the heaviest molecules interfered so far, 
and will still be comparable to superpositions achievable with state-of-the-art optomechanical systems in the near future.

According to Leggett's classification of macroscopically distinct superpositions \cite{Leggett2002}, the mentioned empirical macroscopicity \cite{Nimmrichter2013}  complements other measures found in the literature, where the degree of macroscopicity is associated to entanglement in a many-body superposition state \cite{Bjoerk2004,Korsbakken2007,Marquardt2008,Lee2011,Froewis2012}. For a fixed atom number, a cat, GHZ, or NOON state marks the top end of the yardstick in these approaches  whereas product states are at the microscopic bottom end. Realistic scenarios of weakly interacting two-mode condensates with a controllable amount of squeezing would be somewhere in between, even though the actual entanglement is inferred only indirectly through variances of collective observables, such as the phase and the population difference \cite{Sorensen2001,Hyllus2012,Dalton2014,Lucke2014}.
The interference of Bose-condensed atoms thus brings about  both aspects of macroscopicity: many-body entanglement and sensitivity to MMM.

In a two-mode BEC interferometer, three basic types of nonclassical behaviour have been observed: (i) the bosonic character of the atoms that leads to interference fringes between independent condensates with no phase relation \cite{BECInterferenceKetterle1997}, (ii) phase-stable single-atom interference in a coherently split condensate over spatial \cite{Shin2004,Jo2007,Baumgaertner2010,Schmiedmayer2013,Schmiedmayer2014,CroninReview2009} or internal \cite{Treutlein2004,bohi2009coherent,Gross2010} degrees of freedom, and (iii) genuine many-particle nonlocality by violating Tura-Bell inequalities with number-squeezed condensates \cite{Tura2014,TuraBell2016}. The very first BEC superposition experiments verified indistinguishability (i), possibly the weakest quantum phenomenon in terms of macroscopicity, known also from laser interferometry \cite{Hariharan2007}. 
Phase-coherent interference (ii) and nonlocality tests (iii), on the other hand, are more suited to test macrorealism, as their measurement results would be affected by MMM-induced dephasing at least on the single-particle level.

\begin{figure}
  \centering
  \includegraphics[width=1\textwidth]{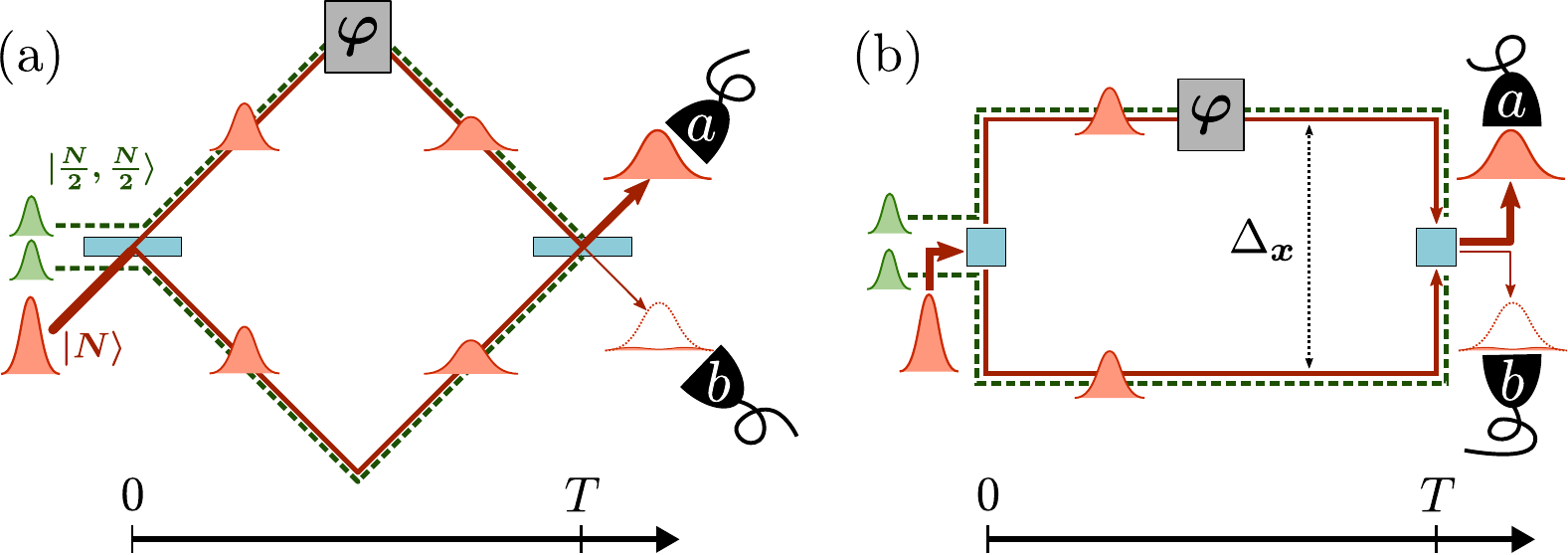}
  \caption{Generic two-arm Mach-Zehnder configuration leading to matter-wave interference in two output modes. (a) $N$ Bose-condensed atoms (red) are coherently split into a superposition of two modes with distinct momenta, evolve dispersively along spatially separate arms, get reflected, and recombined after accumulating a total phase difference $\varphi$ (including the beam-splitter phase). 
By varying  $\varphi$ an interference signal can be recorded in the atom counts detected in each output mode $a,b$. Dephasing and uncontrolled phase fluctuations over many runs result in a washed-out interferogram. Second order interference with a random phase $\varphi$ in each run is also observed by preparing two independent condensates of $N/2$ atoms in each Mach-Zehnder arm (green, dashed). (b) Abstraction of the scheme where the two diverging and reconverging modes  are substituted by two wave packets at rest, displaced by an effective distance $\Delta_x$. Both schemes are essentially equivalent regarding the effect of decoherence and objective collapse.}
\label{PSvsDFS}
\end{figure}

The goal of the present article is to assess the empirical macroscopicity and to discuss the implications of macrorealistic modifications acting on various states of a BEC 
evolving in the two spatially separated arms of a  Mach-Zehnder interferometer, as sketched in Fig.~\ref{PSvsDFS}. 
Our results will also describe the sensitivity of interfering condensate states and their observed signatures to environmental decoherence. In fact, the generic master equation for macrorealistic collapse models adopted here applies to any process that induces single-particle diffusion and dephasing in an exchange-symmetric manner.

For the purpose of evaluating the results, we will focus on a recent experiment with Rb condensates and a vertical arm separation of half a meter \cite{Kovachy2015}. 
The claim based on Extended Data Figure 3 in Ref.~\cite{Kovachy2015} that the measurement rules out MMM at this size scale has caused some controversy \cite{BerkeleyComment,KovachyReply} due to the fact that phase-stable interference fringes in the atom count statistics of the two output ports were not directly observed. 
(Phase stability was improved in a subsequent dual-interferometer scheme at smaller wave packet separation \cite{Asenbaum2016}.) 
According to the criticism, the original experiment leaves room for doubts as to whether a coherent splitting was at all realized in the experiment or whether the same statistics could in principle result from two condensates without phase relation. 
We will show that, from a strictly empirical standpoint, this ambiguity implies a significantly different sensitivity to macrorealistic collapse and decoherence. 

Specifically, we will analyze the impact of MMM on the measurement statistics and results of generic observables for three different condensate states. The empirical macroscopicity of an ideal single-atom interferometer can be reproduced with condensates that form product states (PS) of single-atom superpositions, but not with phase-averaged product states (PAPS) or dual Fock states (DFS). 
Moreover, phase coherence in a single run cannot be deduced from the statistics of atom counts in the two output modes accumulated over many phase-randomized runs. 
This all implies that a PS yields the highest value of macroscopicity in a given phase-stable experiment, equal to that of an equivalent single-atom interferometer operated with the same number of particles. 
Given a fixed population difference between the two arms, a PAPS can be distinguished from a DFS of independent condensates if single atom counts are resolved \cite{Laloe2012}. This level of precision would then imply that also higher-order correlations between the atoms could be detected, which would increase the macroscopicity of the experiment as they are more vulnerable to decoherence.

The paper is organized as follows. In Sec.~\ref{sec:results}, we first introduce the experimental setting and the investigated MMM class of decoherence models, before we present the main results applied to the  atom fountain experiment of Ref.~\cite{Kovachy2015}. Section~\ref{sec:theory} proceeds with a detailed treatment of the predictions of MMM in a single-particle phase-space framework. Using the  formalism of second quantization we then show in Sec.~\ref{sec:2ndQ} how the results of the previous section can be applied to single-atom observables in two-mode condensates and, in typical scenarios where single-mode dispersion is negligible, also to higher-order observables and the full atom count distribution. We conclude with an outlook in Sec.~\ref{sec:concl}.

\section{Two-mode interference, decoherence, and macroscopicity}\label{sec:results}

According to  the empirical measure of macroscopicity \cite{Nimmrichter2013} a BEC interference experiment is to be graded only according to the measurement data. We shall therefore not ask how macroscopic a presumably realized many-body superposition state would be in itself, e.g.~in terms of its entanglement properties. Rather, we assess a quantum experiment by analyzing to what extent it rules out generic decoherence and collapse effects, as associated with a minimal macrorealistic modification (MMM).

The observable consequences of MMM can be described by a universal Lindblad superoperator $\mathcal{L}$ that is added to the von Neumann equation for an arbitrary mechanical system, $\partial_t\rho=\left[\op{H},\rho\right] / i\hbar + \cL \rho$. For the case of a single bosonic particle species of mass $m$ considered here, the  incoherent term can be written in second quantization as \cite{Nimmrichter2013,Nimmrichter2014a}
\begin{eqnarray}
\fl \cL \rho &=&
\frac{m^2}{\tau_e m_e^2 }\int \frac{\diff \vect{r} \diff\vect{r}'\diff\vect{s}}{\left( \sqrt{2\pi} \tilde{\sigma}_s \right)^3} e^{-(\vect{r}-\vect{r}')^2\sigma_q^2/2\hbar^2-s^2/2\tilde{\sigma}_s^2}
\left[\hat{\psi}^\dagger (\vect{r}-\vect{s}) \hat{\psi} (\vect{r})\rho\hat{\psi}^\dagger (\vect{r}')\hat{\psi} (\vect{r}'-\vect{s}) 
\right.\nonumber
\\
\fl && \left.-\frac{1}{2}\hat{\psi}^\dagger (\vect{r}')\hat{\psi} (\vect{r}'-\vect{s}) \hat{\psi}^\dagger (\vect{r}-\vect{s}) \hat{\psi} (\vect{r})\rho
-\frac{1}{2}\rho\hat{\psi}^\dagger (\vect{r}')\hat{\psi} (\vect{r}'-\vect{s}) \hat{\psi}^\dagger (\vect{r}-\vect{s}) \hat{\psi} (\vect{r})
\right]\,.
\label{eq:MMM2ndfull}
\end{eqnarray}
Here, $\hat{\psi} (\vect{r})$ denotes the bosonic field operator; the formulation for a single particle in first quantization follows by replacing $\hat{\psi}^\dagger (\vect{r}) \to |\vect{r}\rangle$. 

Equation (\ref{eq:MMM2ndfull})  describes a gradual decay of motional coherence
with a single-particle rate $1/\tau = (m/m_e)^2/\tau_e$  beyond a  critical length scale of $\hbar / \sigma_q $ and a critical momentum scale of $\hbar/\tilde{\sigma}_s$. 
Its form follows essentially by imposing a number of basic symmetry and consistency requirements 
such as Galilean covariance, exchange symmetry, scale invariance, and boundedness  \cite{Nimmrichter2013}. The positive parameters $\tau_e$, $\sigma_q$, and $\tilde{\sigma}_s$, defined at the reference mass $m_e$ of an electron, are undetermined, but one assumes $\tilde{\sigma}_s< (m_e/m) \times 20\,$pm and $\hbar/\sigma_q<10\,$fm to avoid nuclear excitations.
 A successful quantum superposition experiment provides further bounds on these values,
and the macroscopicity reached can be assessed in terms of the greatest value of $\tau$ ruled out by the observation \cite{Nimmrichter2013}. 
One prominent example of a minimal macrorealistic modification covered by \eqref{eq:MMM2ndfull} is the model of Continuous Spontaneous Localization \cite{Ghirardi1990,Bassi2013}, where $\hbar/ \sqrt{2} \sigma_q=100\,$nm and $\tilde{\sigma}_s=0$. A regularized version of the self-gravitational collapse model~\cite{Diosi1987,Diosi1989} can be brought into a similar form \cite{Nimmrichter2014,Nimmrichter2015}.

We note that for  the dilute and at most weakly interacting atoms considered in this article a finite value of $\tilde{\sigma}_s \ll 10\,$fm  is practically unobservable so that we can safely assume $\tilde{\sigma}_s=0$ in the following.
The dynamics described by Eq.~\eqref{eq:MMM2ndfull}  can then be interpreted as resulting from a point process of random momentum displacements at an average strength $\sigma_q$ and rate $1/\tau$ that does not distinguish between the atoms in the condensate. 

To be explicit, we can rewrite \eqref{eq:MMM2ndfull} with  $\tilde{\sigma}_s=0$ and $\tau = (m_e/m)^2\tau_e$ as 
\begin{equation}
\cL (\rho) = \frac{1}{2\tau} \int \diff \vect{q}\, g(\vect{q}) \left[ \op{A}_{\vect{q}}, \left[ \rho, \op{A}_{\vect{q}}^\dagger \right] \right], 
\label{eq:MMM2nd}
\end{equation}
with $g(\vect{q}) = g(|\vect{q}|)$ an isotropic Gaussian distribution of standard deviation $\sigma_q$. The operator
\begin{equation}
 \op{A}_{\vect{q}} = \int \diff \vect{p}\, \op{a}^\dagger (\vect{p}+\vect{q}) \op{a} (\vect{p}) \,.
\label{eq:MMM2nd_Aq}
\end{equation}
describes a single-particle momentum displacement by $\vect{q}$.
Master equations of this type appear in the description of environmental decoherence, e.g.~by scattering background gas or thermal radiation \cite{gallis1990environmental,Hornberger2003Collisional,hornberger2003collisional2,Vacchini2005,vacchini2007precise}. There, the distribution of momentum displacements is not Gaussian, but determined by the differential scattering cross-section and the ensemble state of the incident scatterers. For instance, the generator \eqref{eq:MMM2nd} could describe decoherence by isotropic scattering of radiation at the atom cloud in each arm, if $g/\tau$ were given by a frequency-dependent scattering rate. Hence, as we analyze the impact of MMM on BEC interferometry in the following, we also make a statement about the sensitivity of many-body superposition states to environmental decoherence.

In the two-mode BEC scenario with well separated arms considered here, MMM induce two effects: (i) decoherence of coherent superpositions at a rate of at most $1/\tau$ and (ii) particle loss from the condensate due to isotropic diffusion heating at $3 \sigma_q^2 /2m\tau $ of power per atom. The macroscopicity of a given condensate state realized in the experiment will thus depend on how sensitive its observed signatures are to those two effects.

In this article, we consider three relevant states of $N$ condensed atoms 
distributed equally into two arms of a Mach-Zehnder setup, as represented by the bosonic annihilation operators $\oc_{a}$, $\oc_{b}$. The first is a product state (PS) of $N$ single-particle superpositions of the form $\psi(\vect{r},t) = \left[ \psi_a (\vect{r},t) + e^{i\phi} \psi_b (\vect{r},t) \right]/\sqrt{2} $, which represents a coherently split BEC with a stable relative phase $\phi$, 
\begin{equation}
|\Psi_{\rm PS} (\phi) \ra = \frac{1}{\sqrt{2^N N!}} \left( \oc_a^\dagger + e^{i\phi} \oc_b^\dagger \right)^N | {\rm vac} \ra. \label{eq:PS}
\end{equation}
It can also be understood as the analog of a coherent state in the language of collective spins \cite{Arecchi1972,Zhang1990,Ma2011}. 
Note that the single-mode wavefunctions $\psi_{a,b} (\vect{r},t)$ may depend explicitly  on time to reflect the free motion and dispersion in each arm. An experimental realization of the PS would result in stable interference fringes as a function of $\phi$, but it requires perfect phase stability from shot to shot. 

For $N\gg 1$, the single-shot behavior then matches the average over many runs, and it can be approximated by the macroscopic wavefunction, an order parameter determining the mean-field density and phase of the condensate \cite{pitaevskii2016bose}. In this case, the MMM-induced dephasing effect can also be accounted for in the mean-field picture, as  discussed in detail in Sect.~\ref{sec:macroWf}. The results are the same as for a single atom: MMM-induced dephasing reduces the average interference fringe visibility ${\cal V} \propto \left| \qerw{ \oc_a^\da \oc_b } \right| $  approximately by the factor
\begin{equation}
D (\sigma_q,\tau) = \exp\left[ -\frac{T}{\tau} \left( 1 - e^{-\Delta_x^2 \sigma_q^2 / 2\hbar^2} \right) \right], \label{eq:D_MMM}
\end{equation}
given the interference time $T$ and the arm separation $\Delta_x$. 

Note that although $D(\sigma_q,\tau)$ can be used to reproduce the green curves of Extended Data Figure 3 in \cite{Kovachy2015}, 
it strictly applies only to the \emph{average} interference contrast. And the observation of an average contrast ${\cal V}_{\rm obs} > 0$ requires a pure PS, or at a best a mildly phase-averaged one. This would then rule out all MMM parameters for which $D(\sigma_q,\tau) < {\cal V}_{\rm obs}$. In particular, those MMM with critical length scales smaller than the arm separation, $\hbar/\sigma_q \lesssim \Delta_x$, are ruled out  most effectively, which yields a macroscopicity 
\begin{equation}
\mu = \log_{10} \left[ \left| \frac{1}{\ln {\cal V}_{\rm obs}} \right| \left( \frac{m}{m_e} \right)^2 \frac{T}{1\,{\rm s}} \right],
\end{equation}
according to the logarithmic measure defined in \cite{Nimmrichter2013}. It would amount to $\mu\approx 12$ for the settings of \cite{Kovachy2015} if a phase-stable interferogram had been recorded at 95\% visibility~\footnote{Notice the weak divergence as ${\cal V}_{\rm obs} \to 1$. Although the macroscopicity grows the closer one gets to a perfect 100\% contrast, measurements can never reach it with certainty, but only up to a finite confidence with lower bound ${\cal V}_{\rm obs} < 1$.}.

The second type of state describes a scenario where the relative phase of the two arms is not stable, but fluctuates randomly from shot to shot, e.g.~due to setup vibrations, as is the case in \cite{Kovachy2015}.  Assuming that the atom number $N$ and the splitting ratio of the condensate remains stable, the measurement statistics accumulated over many shots is described by a phase-averaged ensemble of product states (PAPS),
\begin{equation}
\rho_{\rm PAPS} = \frac{1}{2\pi} \int_0^{2\pi} {\rm d} \phi \, |\Psi_{\rm PS} (\phi) \ra \la \Psi_{\rm PS} (\phi) |. \label{eq:PAPS}
\end{equation}
This state does not exhibit any first-order phase coherence or interference fringes, $\qerw{\oc_a^\da \oc_b} = 0$. Hence it is insensitive to dephasing between the two arms, and MMM affect it only in as much as they deplete the condensate through diffusion heating. 
Yet, in each single run a condensate is coherently split, a fixed (random) phase is established between the arms, and the two parts can interfere upon recombination in the output beam splitter. In fact, a spatial image of the overlapping clouds would reveal interference fringes. 
Theoretically, such random-phase single-shot fringe patterns are reflected in the second-order correlation functions. In the binary setting discussed here, only  the number of atoms  in the two output ports is recorded  per shot; there is no spatial imaging of single-shot interferograms. 
The second-order correlations then describe bunching and anti-bunching, as reflected in the values of $\qerw{ \oc_a^\da \oc_a^\da \oc_a \oc_a}$, $\qerw{ \oc_b^\da \oc_b^\da \oc_b \oc_b}$, and $\qerw{ \oc_a^\da \oc_b^\da \oc_a \oc_b}$. 

The third type of input state considered here, is the dual Fock state (DFS), a pure two-mode state with no phase information,
\begin{equation}
|\Psi_{\rm DFS} \ra = \frac{1}{(N/2)!} \left( \oc_a^\dagger \oc_b^\dagger \right)^{N/2} | {\rm vac} \ra. \label{eq:DFS}
\end{equation}
It represents two \textit{independent} condensates of $N/2$ atoms, each occupying one arm, as was realized in the very first BEC interference experiment \cite{BECInterferenceKetterle1997}, and also in a two-mode setting with twin-atom beams \cite{Lopes2015}. 
Like the PAPS, the DFS yields spatial images with high-contrast interference patterns in every single shot, but with a random phase so that no pattern remains after averaging \cite{Naraschewski1996}. It is thus not affected by MMM-induced dephasing either. The fringes are a consequence of particle exchange symmetry---in itself a quantum feature of atoms---bearing close analogy to laser interferometry \cite{ou2007multi}. 
Both PAPS and DFS are less sensitive to MMM than a PS, i.e.~have lower macroscopicity. This sensitivity does not depend on the arm separation, no matter how  large the latter may be, given a fixed interference time. 

\begin{figure}
  \centering
  \includegraphics[width=.75\textwidth]{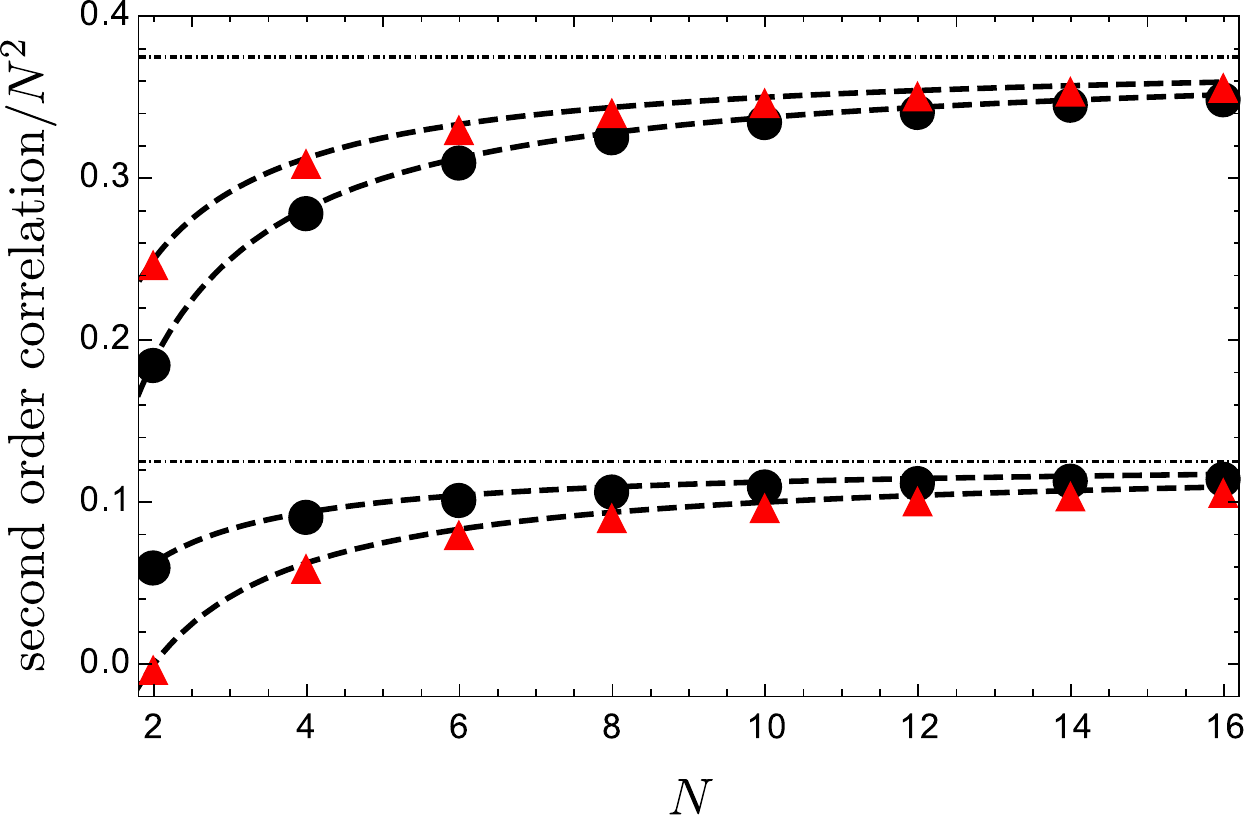}
  \caption{Normalized second order correlation functions for a PAPS (circles) and a DFS (triangles) as a function of the atom number $N$. We assume equal splitting at even $N$. The upper and the lower data  refer to the number of atoms detected in the same and in different modes, respectively. The dashed curves are plotted to guide the eye, and the dash-dotted lines correspond to the asymptotes $N\to\infty$, see Eqs.~(\ref{Corr1PAPS})-(\ref{Corr2DFS}) in Sect.~\ref{sec:dephasing2}. }
\label{secondordercorrelation}
\end{figure}

If the interferometric setup cannot maintain a stable phase it is now natural to ask if there is any way to infer from the measurement data whether a coherently split condensate or two separate ones were present in each run? In other words, do the experimental signatures of a PAPS and a DFS differ? Our detailed assessment in Sect.~\ref{sec:2ndQ} shows that these states can be distinguished, but it requires a high precision, down to the level of single atoms, both in the preparation of the condensate and in the count statistics.

This is exemplified by Fig.~\ref{secondordercorrelation}, which compares  second-order correlations of a PAPS (circles) to those of a DFS (triangles) as a function of the atom number $N$. The upper and the lower data sets correspond to the normalized normally ordered products of the atom count number in the same and in different output ports, respectively, i.e., $\qerw{ \oc_a^\da \oc_a^\da \oc_a \oc_a}/N^2$ and $\qerw{ \oc_a^\da \oc_b^\da \oc_a \oc_b}/N^2$. 
We observe that the correlations for PAPS and DFS deviate if the total number of atoms is small, but the  difference decreases as the asymptotic bounds (grey lines) are approached with growing $N$.

The observable difference between a PAPS and a DFS is most clearly captured by the full atom count statistics in the output ports of the final beam splitter, 
as discussed in Sect.~\ref{sec:measStatistics}. At even $N$, the DFS yields alternating counting probabilities $P (n)$ with destructive Hong-Ou-Mandel interference at odd numbers and constructive interference at even numbers, whereas the PAPS probabilities approximate the classical continuous distribution $g(p) = 1/\pi \sqrt{p(1-p)}$ for $p=n/N$ and $N\gg 1$. 
If the alternating behavior of adjacent count numbers $n$ and $n\pm 1$ cannot  be resolved due to a finite measurement resolution a  DFS cannot be distinguished from a PAPS. 

\begin{figure}
  \centering
  \includegraphics[width=.75\textwidth]{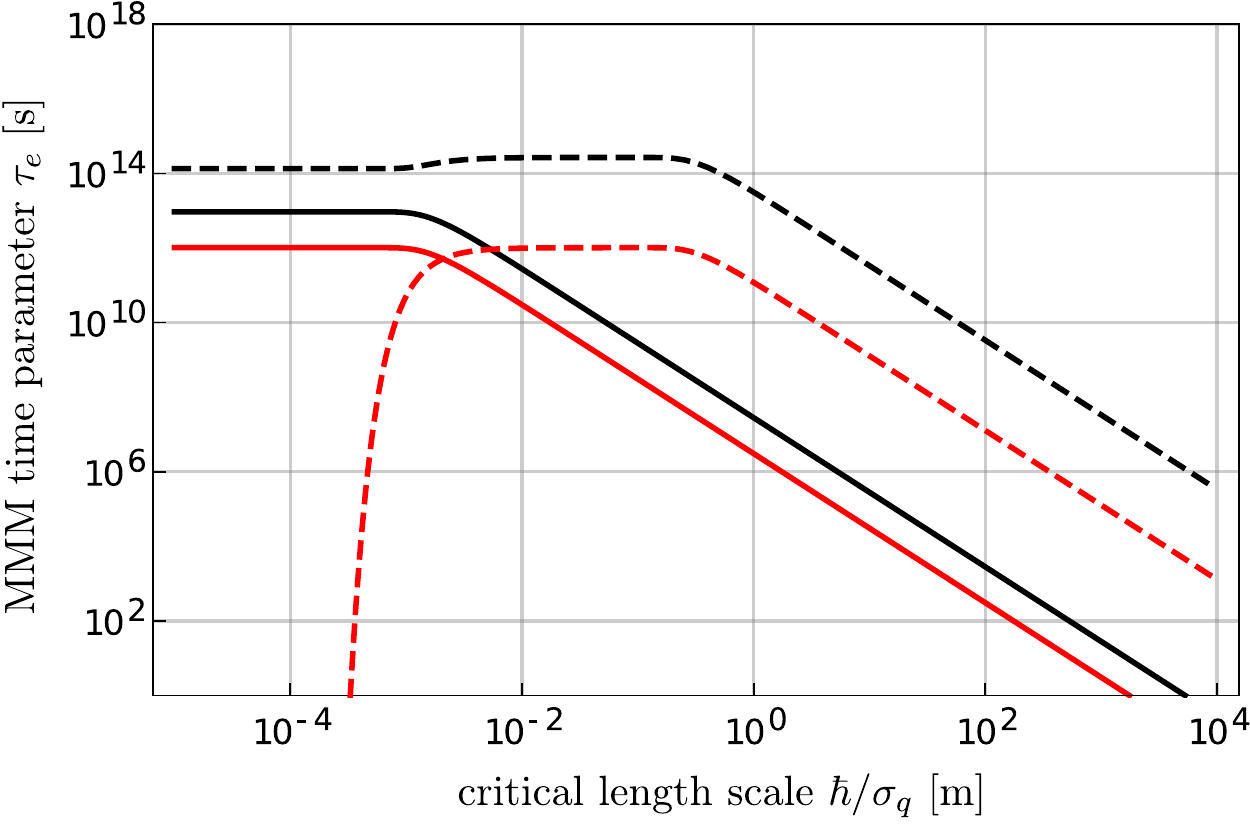}\\
  \caption{MMM parameters excluded by various realizations of Mach-Zehnder interference with ${}^{87}$Rb condensates, assuming $T=2.08\,{\rm s}$ interference time, $\Delta_x=0.5\,{\rm m}$ arm separation, and $w_{x,y,z}=1\,{\rm mm}$ initial width of the condensate. Detecting an average of at least $95\%$ of the initially prepared atoms in the output ports rules out the region below the red solid line, while measuring an average $95\%$ of phase-stable interference contrast excludes everything below the red dashed curve. 
  The black solid line corresponds to measuring a specific many-body observable (\ref{eq:HOMobservable}) at 80~\% accuracy that captures the Houng-Ou-Mandel signature of two independent condensates.
  The black dashed curve represents an experimental sample of the atom count distribution $P(n_a)$ in one of the output ports, where the variance in $n_a$ is no more than 40 times the ideal value predicted for a coherently split condensate.}
\label{fig:exclusionplots}
\end{figure}

Figure~\ref{fig:exclusionplots} compares the three types of states in terms of their sensitivity to MMM decoherence in the experiment. It shows the range of MMM parameters $\tau_e, \sigma_q$ ruled out by various (hypothetical) measurement scenarios in a Mach-Zehnder setup with Rb condensates, assuming the parameters of Ref.~\cite{Kovachy2015}. The red solid line corresponds to the experimental observation of less than 5~\% atom loss during the interference time $T$. All MMM parameters below the curve are then excluded by the measurement data as they would cause a stronger depletion of the condensate. The curve assumes a maximum where the critical MMM length scale $\hbar/\sigma_q$ is smaller or comparable to the dimension of the atom cloud in each arm. The measurement does not require any phase coherence, any condensate state would give the same result. In fact, one could achieve the same sensitivity by detecting atom losses in a single-mode condensate. 

The black solid line in Fig.~\ref{fig:exclusionplots} represents the sensitivity of Hong-Ou-Mandel (HOM) interference with a DFS of 30 atoms, assuming that the particular weighted HOM dip visibility (\ref{eq:HOMobservable}) defined in Sec.~\ref{sec:measStatistics} can be extracted from the count statistics at more than 80~\% of its ideal value. This scenario, which requires single-atom precision in the state preparation and detection, rules out more parameters than the depletion measurement, but the relevant length scale is still set by the mm-size of the atom cloud. A high sensitivity on the half-metre scale is only achieved with PS, i.e.~in experimental scenarios with coherently-split condensates and phase-stable interferograms. 
The red dashed curve corresponds to an observation of phase-stable interference fringes of more than 95~\% contrast. Here, the sensitivity drops sharply in the regime where MMM would predominantly deplete the condensate since the interference signal comprises only the remaining atoms. 

Finally, the black dashed curve in Fig.~\ref{fig:exclusionplots}  showcases what could be achieved by sampling over many runs of phase-stable coherent splitting; it assumes that the atom count distribution of a PS of $10^5$ atoms at a fixed relative phase is recorded in one output port with a measurement accuracy in the atom number of 1~\%. This specific experiment could resolve an increase of the initial variance in the atom number by more than a factor 40, thereby ruling out the widest range of MMM parameters. It yields the highest macroscopicity of the four plotted examples, which implies that it is also the most sensitive to environmental decoherence. 
A quantitative derivation of the results in Fig.~\ref{fig:exclusionplots} will be given in Sec.~\ref{sec:2ndQ}.

All three types of states discussed here have in common that they contain no useful form of many-body entanglement (apart from what appears to be entangled in the first quantization picture after symmetrization). Hence once might expect that their realization in the lab cannot be more sensitive to MMM decoherence than an equivalent single-atom interferometer at a comparable number of repetitions. We will see in the following that this holds true as long as single-atom observables are measured. Precision experiments detecting higher-order observables, however, can lead to an enhanced measure of macroscopicity. Whether and how experimental signatures of genuine entanglement, such as squeezing, can go beyond the single-particle level in terms of macroscopicity is a different issue that will be discussed elsewhere.

\section{Theoretical model of the MMM
effect}\label{sec:theory}

In the following, we describe in detail a two-mode model for the BEC Mach-Zehnder interferometer and the influence of minimal macrorealistic modifications and MMM-like decoherence. We start with a few assumptions that will simplify the calculations, before we discuss the impact of MMM on the single-particle level in Sect.~\ref{sec:macroWf}. 
The results will be employed in Sect.~\ref{sec:2ndQ} to analyze the MMM effect in a second quantization picture.

\subsection{Basic assumptions}\label{sec:assumptions}

The main ingredients for our theoretical description of the two-mode interferometer sketched in Fig.~\ref{PSvsDFS} are the wavefunctions $\psi_{a,b}$ representing the two Mach-Zehnder arms. We assume them to be one-dimensional Gaussian wave packets of equal initial width $w_x$. The arm splitting shall occur in the form of a momentum splitting $\Delta_p$ along the $x$-coordinate, and we will  ignore the condensate profile along $y,z$ most of the time. It will play a role in the assessment of MMM-induced heating depletion later. Moreover, we assume that the wave packets are approximately orthogonal and clearly separated for almost the entire interference time $T$, i.e., $\Delta_p T /m \gg w_x$. 

Immediately after the first beam splitter the configuration of Fig.~\ref{PSvsDFS}a corresponds to a superposition of wave packets that are separated by $\Delta_p$ in momentum,
$\psi_a (x) = (2\pi w_x^2)^{-1/4} \exp ( -x^2/4w_x^2 )$ and $\psi_b (x) = \exp (i\Delta_p x / \hbar) \psi_a (x)$.
Note that, without loss of generality, we operate in the rest frame of arm $a$. The MMM 
are invariant under Galilean transformations between inertial frames, and the constant gravitational acceleration in a vertical setup can be accounted for in the relative phase $\phi$. In order to recombine the wave packets at the second beam splitter at time $T$, the motion of arm $b$ shall be reversed at $T/2$ with help of a momentum displacement operation by $-2\Delta_p$.

As shown in \ref{app:MomentumSplit}, we can increase the level of abstraction by modeling the two-mode configuration as a superposition of spatially separated wave packets at rest, i.e.~$\psi_b (x) = \exp (i\phi) \psi_a (x+\Delta_x) $ with an effective relative phase $\phi$ and an average arm separation $\Delta_x = \Delta_p T/(2\sqrt{3} m)$.
The time evolution then reduces to a free dispersion described by the complex  time-dependent term $w_x^2 (t) = w_x^2(1 + i \hbar t / 2 m w_x^2)$. As far as the MMM-induced decoherence effect is concerned, this is a good approximation provided that $\Delta_x \gg |w_x (t)|$ at all times.

Now let us associate Fock space annihilation operators $\oc_{a,b}$ with the two dispersing modes at time $t$. Expanded in terms of the Schr\"odinger picture field operators, we have 
\begin{equation}
\oc_{a}^\da (t) = \sqrt{\frac{w_x}{\sqrt{2\pi} w_x^2 (t)}}\int \diff x\, e^{-x^2/4w_x^2(t)} \hat{\psi}^\da (x), \label{eq:ca_t}
\end{equation}
and $\oc_{b}^\da (t)$ defined accordingly using the $\Delta_x$-shifted wave packet. 
In the absence of MMM,
the  Heisenberg picture renders these  mode operators constant, since the unitary evolution of the  field operators cancels the explicit time dependence in Eq.~(\ref{eq:ca_t}). To account for MMM we will switch to the interaction picture below where the  time-dependence of the $\oc_{a,b}$ describes the interplay of wave packet dispersion and incoherent dynamics.

The recombining beam splitter is represented by the linear transformation
\begin{eqnarray}
\oc^{\da}_{a,{\rm out}} &=&
\cos(\theta)\oc^{\da}_{a} (T) - \sin(\theta)e^{i\alpha}\oc^{\da}_{b} (T)
,\nonumber \\
\oc^{\da}_{b,{\rm out}} &=&
\sin(\theta)e^{-i\alpha}\oc^{\da}_{a} (T) + \cos(\theta)\oc^{\da}_{b} (T), \label{eq:cab_out}
\end{eqnarray}
with $\alpha$ a tunable phase shift between the transmitted and the reflected state. It can be absorbed in the overall phase between the arms, $\phi \to \varphi = \phi-\alpha$. We assume a balanced splitting into the output ports, $\theta = \pi/4$. The atom count statistics are then determined by the respective number operators and their eigenstates. 

To account for the effect of MMM we can distinguish between two limiting regimes, depending on the average strength $\sigma_q$ of the momentum displacements: (i) $\sigma_q \gg \hbar/w_x$, where MMM are equivalent to a random loss process depleting the condensate at the rate $1/\tau$, and (ii) $\sigma_q \ll \hbar/w_x$, where the MMM-induced dephasing dominates  the atom loss caused by diffusion. Notice here that the assumption of perfect mode matching between the beam splitter transformation and the dispersed wave packets implies that we overestimate the MMM-induced atom loss---a consequence of the reduced two-mode description. 
In principle, MMM diffusion in (ii) causes an additional incoherent broadening of the condensate in each arm, and a realistic beam splitter would transform 
both the condensate and the thermal cloud, essentially without losses. In other words, if the overlapping clouds were directly imaged, the result would be washed out fringes instead of a lower atom number. To estimate the incoherent broadening in the present two-mode setting, one may consider the worst case scenario: all atoms lost from the two-mode condensate state remain in the cloud that arrives at the recombining beam splitter and form a phase-incoherent thermal background that enters both output ports and reduces the interference contrast. For an even more realistic assessment, the exact beam splitter profile must be taken into account to determine the fraction of incoherent background contributing to the detection signal in a particular experimental setting. Here we neglect this contribution and resort to the best case scenario that none of the atoms lost from the condensate will be detected.

\subsection{Effective single-particle treatment}\label{sec:macroWf}

If a single atom travels through the interferometer  in the presence of MMM the  time evolution can be solved analytically in the characteristic function representation \cite{Nimmrichter2013,Nimmrichter2014a}. Accordingly, this solution applies  as well to the time evolution of single-particle operators in second quantization if we introduce the single-particle characteristic function operator,
\begin{equation}
\ochi (x,p) = \int\diff x_0 \, e^{ipx_0/\hbar} \opsi^\da \left( x_0 + \frac{x}{2} \right) \opsi \left( x_0 - \frac{x}{2} \right).\label{eq:chi2nd}
\end{equation}
In the same way as every rank-one operator $\oM_{ij} = |\psi_i \ra \la \psi_j|$ in first quantization is fully  characterized by its characteristic function, we can expand normally ordered operators of the form $\oc^\da_i \oc_j$  in terms of \eqref{eq:chi2nd}. To be specific, 
\begin{equation}
\oc^\da_i \oc_j = \int \frac{\diff x \diff p}{2\pi\hbar} M_{ij} (-x,-p) \ochi (x,p) \label{eq:cicj_chiexpand} ,
\end{equation}
with $M_{ij}(x,p)$ the characteristic function representation of the operator $\oM_{ij} = |\psi_i \ra \la \psi_j|$; for projectors we use the convention $M_{ii}(x,p)=P_i(x,p)$.
The operator $\ochi(x,p)$ constitutes an eigenvector of the  MMM generator \eqref{eq:MMM2nd} (in its 1D version),
\begin{equation}
\cL \left[ \ochi (x,p) \right] = \frac{e^{-x^2\sigma_q^2/2\hbar^2} - 1}{\tau} \ochi (x,p). \label{eq:MMMchi2nd}
\end{equation}
This facilitates explicit solutions of the non-interacting many-body problem in presence of MMM. 

The multiplicative term in (\ref{eq:MMMchi2nd}) and the expansion \eqref{eq:cicj_chiexpand} suggest that the modified time evolution of single-particle expectation values is best described in terms of effective equations of motion in the first-quantization framework. In the end, we will then be left with evaluating the expectation values of $\ochi (x,p)$ with respect to the \textit{initial} states (PS, PAPS, DFS) of two-mode condensates. Expanding the field operators in the two-mode basis and omitting all unoccupied mode contributions one obtains
\begin{equation}
\qerwsm{\ochi (x,p)} = P_a (x,p) \qerwsm{\oc_a^\da \oc_a} + P_b (x,p) \qerwsm{\oc_b^\da \oc_b} + M_{ba} (x,p) \qerwsm{\oc_a^\da \oc_b} + M_{ab} (x,p) \qerwsm{\oc_b^\da \oc_a},\label{eq:chi2nd2mode}
\end{equation}
with the characteristic functions given in Eq.~(\ref{eq:PabMab}) below.

At this point, one might ask whether MMM decoherence can be accounted for in the effective mean-field description of condensates by means of a macroscopic order parameter \cite{leggett2006quantum}. In this spirit, Wallis et al.~\cite{Wallis1997} introduced a Wigner function operator in second quantization---the Fourier transform of \eqref{eq:chi2nd}---and suggested to replace the field operators by a macroscopic wavefunction that describes the condensate in a single run of the experiment. In the present case, this would be  $\psi_{\rm mac} (x,t) = \sqrt{N/2} \left[ \psi_a (x,t) + e^{i\phi} \psi_b (x,t) \right]$ \cite{Naraschewski1996,Wallis1997}. Equations~\eqref{eq:chi2nd} and \eqref{eq:MMMchi2nd} imply that this effective mean-field phase-space treatment yields the correct predictions for expectation values of single-particle observables, implying an average over many runs. However, switching from $\psi_{\rm mac}$ to the mean-field Wigner function and subjecting it to incoherent dynamics does not necessarily reflect the time evolution of the condensate in a single run. 

We will stay on the safe side and address random single-run interference only in terms of second-order correlations in the overall measurement statistics, for which the single-particle treatment does not apply directly. 
Moreover, we do not incorporate atom-atom interactions in the form of a Gross-Pitaevskii equation \cite{LeggettGPE2001}, which would lead to a buildup of correlations \cite{Fabrocini1999}, broadening of the condensate \cite{Gotlibovych2014}, and phase dispersion between the two modes \cite{Javanainen1997PhaseDispersion}. By restricting to dilute, non-interacting condensates we neglect these additional effects and attribute a potential loss of interference visibility entirely to MMM. 

\subsubsection{Effective time evolution}

A general solution for the modified time evolution of $N$-atom two-mode condensate states $\rho(t)$ in the presence of MMM is tedious to compute. Alternatively, we can resort to effective equations of motion for the first- and second order correlation functions.
They are the expectation values of normally ordered products of the output modes' creation and annihilation operators \eqref{eq:cab_out}, whose counterparts in first quantization are linear combinations of the two projectors $\oP_{a,b} (t) = |\psi_{a,b} (t) \ra \la \psi_{a,b} (t) |$, and of the operator $\oM_{ab} (t) = |\psi_{a} (t) \ra \la \psi_{b} (t) |$ and its conjugate at $t=T$. 

In the Schr\"{o}dinger picture used so far, the expectation values 
with respect to the two wave packets traveling through the interferometer 
are determined by a time-dependent state $\rho(t)$ of $N$ atoms.
We now focus on the MMM effects by switching into the interaction picture. 
Of course, it is not guaranteed that the MMM master equation retains a simple form in the interaction frame, but it turns out that it does so for single-particle states and the corresponding observables.

In the interaction picture in first quantization a single-particle state $\rho_I (t) = \oU^\da (t) \rho (t) \oU(t) $, with $\oU(t) = \exp (-i\op{p}^2 t/2m\hbar )$, evolves under MMM according to 
\begin{equation}
\partial_t \rho_I (t) = \frac{1}{\tau} \left[ \int \frac{e^{-q^2/2\sigma_q^2} \diff q}{\sqrt{2\pi}\sigma_q } e^{-i q (\ox+\op{p} t/m)/\hbar} \rho_I (t) e^{i q (\ox+\op{p} t/m)/\hbar} -\rho_I (t) \right]. \label{eq:MMM1st_I}
\end{equation}
Once again, this master equation has an explicit solution in the characteristic function representation. For the evolution of expectation values of an observable  $\oO $, one may as well solve the adjoint equation,  given by the same Lindblad generator. We define its characteristic function representation as
\begin{equation}
O_I(x,p) = \int\diff x_0 \, e^{i p x_0 / \hbar} \la x_0 - \frac{x}{2} | \oO_I | x_0 + \frac{x}{2} \ra.\label{eq:charfunction}
\end{equation}
As discussed in Sect.~\ref{sec:assumptions}, in the absence of MMM the relevant observables are constant in the interaction frame.
In presence of MMM 
the adjoint solution to \eqref{eq:MMM1st_I} then results in a modified characteristic function $O_I (x,p,t)$ that explicitly depends on time; it is multiplied by an exponential decoherence factor, 
\begin{equation}
O_I(x,p) \to O_I(x,p,t) = \exp \left[ \int_0^t \frac{\diff t'}{\tau} e^{-\sigma_q^2 (x+pt'/m)^2/2\hbar^2} - \frac{t}{\tau}  \right] O_I(x,p). \label{eq:OHsol1}
\end{equation}
At this point, we do not carry out the time integration in the exponent (which would result in an error function). 
Expectation values are obtained from the overlap integral $\qerw{\oO(t)} = \int \diff x\diff p\, O_I (-x,-p,t) \chi_0 (x,p) /2\pi\hbar$, with $\chi_0$ the characteristic function of the \textit{initial} state $\rho(0)$.
Here, it is sufficient to consider the characteristic functions
appearing in (\ref{eq:chi2nd2mode}), 
\begin{eqnarray}
\fl P_a (x,p) &=& \exp \left( - \frac{x^2}{8w_x^2} - \frac{p^2w_x^2}{2\hbar^2} \right), \quad P_b(x,p) = P_a (x,p) e^{ i p \Delta_x / \hbar}, \nonumber \\
\fl M_{ab}(x,p) &=& \exp \left[ - \frac{(x-\Delta_x)^2}{8w_x^2} - \frac{p^2w_x^2}{2\hbar^2} + \frac{ip\Delta_x}{2\hbar} \right], \quad M_{ba}(x,p) = M_{ab}^{*} (-x,-p). \label{eq:PabMab}
\end{eqnarray}

We proceed by approximating the MMM effect for these operators in the relevant parameter regimes. It will turn out that, for all practical purposes, the dispersion of the two arms can be ignored in the evaluation of MMM-induced heating and dephasing.

\subsubsection{Strong depletion regime}

A simple result follows in the limit of strong MMM-induced momentum displacement, $\sigma_q \gg \hbar/ w_x$. When applied to the Gaussian functions \eqref{eq:PabMab}, the decoherence factor in \eqref{eq:OHsol1} can be approximated by a uniform exponential decay $\exp(-t/\tau)$ almost everywhere, except at $(x,p)\approx 0$ where it assumes unity, 
\begin{eqnarray}
\fl O_I(x,p,t) &\approx& e^{-t/\tau} O_I(x,p) + e^{-t/\tau} \left\{ \exp \left[ \int_0^t \frac{\diff t'}{\tau} e^{-\sigma_q^2 (x+pt'/m)^2/2\hbar^2} \right] -1 \right\} O_I(0,0) \nonumber \\
\fl &\approx& e^{-t/\tau} O_I(x,p). \label{eq:MMM_1P_strong}
\end{eqnarray}
The second term in the first line vanishes for $M_{ab}$, but we also neglect it for $P_{a,b}$ since it will contribute only a small correction suppressed by $\hbar/\sigma_q w_x$ to the overlap integrals of expectation values. Hence, the exponential decay affects both the two-mode occupation and the coherence $M_{ab}$ in approximately the same way. 

This limit describes  an exponential depletion of the two-mode condensate at the rate $1/\tau$. To see this, we identify the number operator for the two arms, $\op{N}(t) = \oc_a^\da (t) \oc_a (t) + \oc_b^\da (t) \oc_b (t)$, as the second-quantization counterpart of the single-particle projector $\oP(t) = \oP_a(t) + \oP_b(t)$. The average atom number at the recombining beam splitter, and thus the mean atom count rate in the output ports, is then reduced to the fraction $\exp(-T/\tau)$. 

At the same time, the expectation value of the interference term $\oc_a^\da (t) \oc_b (t)$ decays with equal rate. This implies that the average first-order interference \textit{contrast} between the output ports, which is proportional to $|\qerw{\oc_a^\da (T) \oc_b (T)}|/\qerw{\op{N} (T)}$, remains unaffected. The reason is that our two-mode model assumes that none of the atoms that are lost from the condensate will ever reach the detectors. As discussed above, an alternative worst-case estimate accounts for the lost fraction $1-\exp(-t/\tau)$ of atoms in terms of a flat background that raises the mean count rate and thereby decreases the interference contrast. An experiment could then rule out more MMM parameters than what the more conservative two-mode model predicts. In practice, the actual MMM-induced loss will depend on how much background can reach the final beam splitter, which would require a full solution of the modified many-body time evolution.

\subsubsection{Dephasing regime}

In the regime of small momentum parameters, $\sigma_q \ll \hbar/ |w_x (t)|$, MMM decoherence cannot `resolve' the size of the individual dispersing wave packets. The depletion rate of the condensate is then suppressed, whereas dephasing between the two arms can still occur at the full MMM rate $1/\tau$ as long as $\sigma_q \gg \hbar/\Delta_x$.

Applying \eqref{eq:OHsol1} to the populations, we can expand to lowest order in the argument of the Gaussian function of the MMM term,
\begin{eqnarray}
P_{a,b}(x,p,t) &\approx& \exp \left[ -\frac{\sigma_q^2 t}{2\hbar^2 \tau} \left( x^2 + \frac{xpt}{m} + \frac{p^2t^2}{3m^2} \right) \right]  P_{a,b}(x,p). \label{eq:MMM_1P_deph_P}
\end{eqnarray}
For the coherence term $M_{ab}$, we may ignore the small variations of the argument around $x=\Delta_x\gg |w_x(t)|$ and find to lowest non-vanishing order 
\begin{equation}
M_{ab}(x,p,t) \approx \exp \left[ -\frac{t}{\tau} \left( 1 - e^{-\sigma_q^2 \Delta_x^2/2\hbar^2} \right) \right]  M_{ab}(x,p). \label{eq:MMM_1P_deph_M}
\end{equation}
Strictly speaking, this expression is only valid for $\sigma_q \ll \hbar / \sqrt{\Delta_x w_x}$, which is a tighter requirement for the dephasing regime. However, since we also assume $\Delta_x \gg w_x$, corrections to the Gaussian in \eqref{eq:MMM_1P_deph_M} will only matter in a regime where the Gaussian is already nearly zero anyway, $\sigma_q \gtrsim \hbar / \sqrt{\Delta_x w_x} \gg \hbar / \Delta_x$. In this limit, MMM dephasing occurs at the full rate $1/\tau$. 

Notice that the decay of $M_{ab}$ in the dephasing regime translates directly into the loss of interference contrast described by \eqref{eq:D_MMM}, $M_{ab} (x,p,T)=D(\sigma_q,\tau)M_{ab}(x,p)$. It also agrees with the result \eqref{eq:MMM_1P_strong} for the strong depletion regime at large $\sigma_q$. The crucial point is that dephasing saturates at the maximum rate if $\sigma_q \gg \hbar / \Delta_x$, whereas strong depletion requires $\sigma_q \gg \hbar / w_x$.

\subsubsection{The relevance of dispersion}

We can now evaluate the expectation values of the population and coherence between the two modes by integrating the modified functions $P_{a,b}(x,p,t)$ and $M_{ab}(x,p,t)$ with the  characteristic function associated to the initial state $\rho_0=\rho_I (0)$. Under the assumption of far separated and practically orthogonal wave packets, $\Delta_x \gg w_x$, we are left with
\begin{eqnarray}
\fl \qerw{\oP_{j=a,b} (T)} &\approx& \la \psi_j |\rho_0 |\psi_j \ra \int \frac{\diff x\diff p}{2\pi\hbar}\, P_j (-x,-p,T) P_j(x,p) \stackrel{\sigma_q \gg \hbar/w_x}{\longrightarrow} e^{-t/\tau} \la \psi_j |\rho_0 |\psi_j \ra, \label{eq:erwP_MMM_1P}\\
\fl \qerw{\oM_{ab} (T)} &\approx& \la \psi_b |\rho_0 |\psi_a \ra \int \frac{\diff x\diff p}{2\pi\hbar}\, M_{ab} (-x,-p,T) M_{ba}(x,p) \approx D(\sigma_q, \tau) \la \psi_b |\rho_0 |\psi_a \ra. \label{eq:erwM_MMM_1P}
\end{eqnarray}
In the parameter regimes discussed here, the size and dispersion of the wave packets  impacts Eq.~(\ref{eq:erwP_MMM_1P}) only in the dephasing regime, 
\begin{eqnarray}
\fl \int \frac{\diff x\diff p}{2\pi\hbar}\, P_j (-x,-p,T) P_j(x,p) \nonumber \\
\approx \left\{ 1 + 2\left( \frac{\sigma_q w_x}{\hbar}\right)^2 \frac{T}{\tau} \left[ 1 + \frac{1}{6} \left( \frac{\hbar T}{mw_x^2} \right)^2 + \frac{T}{24\tau} \left( \frac{\sigma_q T}{mw_x} \right)^2\right] \right\}^{-1/2} .  \label{eq:erwP_MMM_1P_app_disp}
\end{eqnarray}
Here, the higher powers in $T$ are corrections due to the dispersion of the wave packets. They are of second order in the ratio of the interference time $T$ and the characteristic diffraction time scale $t_{\rm d} = mw_x^2 / \hbar$ of a Gaussian wave. For the experimental scenario studied here $T/t_{\rm d} \ll 1$, which renders the free dispersion irrelevant. We may then approximate
\begin{eqnarray}
\fl \int \frac{\diff x\diff p}{2\pi\hbar}\, P_j (-x,-p,T) P_j(x,p) &\approx&  
\exp \left[ -\frac{T}{\tau} \left( 1 - e^{-\sigma_q^2 w_x^2/\hbar^2} \right) \right] \equiv H(\sigma_q,\tau),  \label{eq:erwP_MMM_1P_app}
\end{eqnarray}
which also assumes a not too large $T/\tau$---a technicality since the strong MMM-induced depletion implied at $\tau \ll T$ is in any case ruled out by the experiment.

The depletion term $H(\sigma_q,\tau)$, which determines the probability for an atom to remain in the condensate, is now written in such a way that it interpolates between the dephasing and the strong depletion regime. Both \eqref{eq:MMM_1P_deph_M} and \eqref{eq:erwP_MMM_1P_app} describe these regimes correctly, but in the transient regime of $\sigma_q \sim \hbar/w_x$ they can serve  only as qualitative estimates of the MMM effect. 
It will be an essential prerequisite for obtaining analytically tractable results in the subsequent many-body treatment that we omit the free dispersion. This amounts to removing the free Hamiltonian $\oH_0$ from the time evolution and solving the simplified master equation $\partial_t \rho = \cL \rho$ with two static modes. 

So far, we have described the condensate state in terms of one-dimensional wave functions, assuming that the spatial separation takes place along the $x$-direction. Indeed, the shape and confinement of the condensate in the perpendicular directions $y,z$ does not affect the MMM-induced dephasing between the two arms, as given by \eqref{eq:erwM_MMM_1P}. But the heating rate \eqref{eq:erwP_MMM_1P_app} describing the atom loss in each arm does increase as the atoms are also subject to diffusion along $y$ and $z$. Assuming a common gaussian profile for both arms with waists $w_y,w_z$, we can account for the additional diffusion by replacing the above heating rate with 
\begin{eqnarray}
H(\sigma_q,\tau)=\exp \left[ -\frac{T}{\tau} \left( 1 - e^{-\sigma_q^2 (w_x^2+w_y^2+w_z^2)/\hbar^2} \right) \right], \label{eq:H_3D}
\end{eqnarray}
i.e.~substituting $w_x$ by $w= \sqrt{w_x^2+w_y^2+w_z^2}$. For isotropic wave packets, this triples the effective loss rate compared to the one-dimensional description in the dephasing regime, while the rate remains the same $1/\tau$ in the strong depletion limit.

\section{BEC interference in second quantization}\label{sec:2ndQ}

We now study the impact of MMM on the interference visibility and measurement statistics of the Mach-Zehnder interferometer for the three types of two-mode BEC states  introduced in Sect.~\ref{sec:results}, the PS, the PAPS, and the DFS. 
We first consider expectation values of first-order (or single-particle) observables exhibiting stable interference fringes for phase-coherent PS. We then extend our dispersionless treatment of MMM to the second-order (two-particle) correlations that are necessary to account for the random-phase single-shot interference of PAPS and DFS. Finally, we evaluate the phase-averaged atom count statistics in the output ports and find that it is insensitive to decoherence and affected  only by MMM-induced particle loss.

\subsection{First-order interference}\label{sec:dephasing1}

First-order Mach-Zehnder interference of single atoms or coherently split condensates yields a phase-dependent fringe oscillation of the atom counts in each output port, as reflected in the first-order expectation values $\qerw{\oc_{a,{\rm out}}^\da \oc_{a,{\rm out}}}$ and $\qerw{\oc_{b,{\rm out}}^\da \oc_{b,{\rm out}}}$. By virtue of the beam splitter transformation \eqref{eq:cab_out} and the two-mode characteristic function expansion in \eqref{eq:cicj_chiexpand} and \eqref{eq:chi2nd2mode}, we can apply the results of the previous section and solve the modified evolution of the single-particle observables in the interaction picture. 
Assuming non-overlapping arms, we get 
\begin{eqnarray}
\fl \qerw{\oc_{a,{\rm out}}^\da \oc_{a,{\rm out}}} &=& \qerw{\frac{\oc_a^\da \oc_a + \oc_b^\da \oc_b}{2}} H(\sigma_q,\tau) - \qerw{\frac{e^{i\alpha}\oc_b^\da \oc_a + e^{-i\alpha}\oc_a^\da \oc_b}{2}} D(\sigma_q,\tau), \nonumber \\
\fl \qerw{\oc_{b,{\rm out}}^\da \oc_{b,{\rm out}}} &=& \qerw{\frac{\oc_a^\da \oc_a + \oc_b^\da \oc_b}{2}} H(\sigma_q,\tau) + \qerw{\frac{e^{i\alpha}\oc_b^\da \oc_a + e^{-i\alpha}\oc_a^\da \oc_b}{2}} D(\sigma_q,\tau).
\end{eqnarray}
Here, we evaluate the expectation values with respect to the undispersed modes and the initial condensate states. The first terms involve the atom number operator $\op{N} = \oc_a^\da\oc_a + \oc_b^\da\oc_b$ and represent the average count rate if no phase information is available. The second terms describe first-order interference. We restrict to output port $a$ in the following.

For the phase-coherent PS \eqref{eq:PS}, the fringe pattern appears as a function of the difference $\varphi = \phi - \alpha$ of the relative phase $\phi$ between the two arms and the beam splitter phase $\alpha$,

\begin{equation}
\qerw{\oc_{a,{\rm out}}^\da \oc_{a,{\rm out}}}_{\rm PS} = \frac{N}{2} \left[ H(\sigma_q,\tau) - \cos (\varphi) D(\sigma_q,\tau) \right]. \label{eq:caca_PS}
\end{equation}
The fringe contrast diminishes in proportion to $D(\sigma_q, \tau) / H(\sigma_q, \tau)$. It is well approximated by $D(\sigma_q, \tau)$ in the relevant dephasing regime, as already anticipated in \eqref{eq:D_MMM}. In the strong depletion regime, on the other hand, MMM yield no loss of contrast since all affected atoms are removed from the condensate.

For the PAPS \eqref{eq:PAPS} and the DFS \eqref{eq:DFS}, there is no phase information when averaged over many runs, and the first-order coherence vanishes, $\qerw{\oc_a^\da \oc_b}=0$. The ensemble-averaged count rate in each port is given by $N/2$ times the MMM depletion term $H(\sigma_q,\tau)$. This merely allows one to rule out those MMM parameters $(\sigma_q,\tau)$ that predict a higher overall loss of atoms than observed (provided that the initial atom number is known to some extent). 
For example, if we assume that the PAPS observed in the interferometer of Ref.~\cite{Kovachy2015} still contains a fraction $P \gtrsim 95\%$ of the initially prepared atoms upon detection, this rules out all MMM parameters with $H(\sigma_q, \tau) < P$. Following Ref.~\cite{Nimmrichter2013}, the greatest ruled out $\tau$ parameter, i.e.~the maximum of the function $\tau(\sigma_q)$ given by $H(\sigma_q,\tau) = P$, would set the macroscopicity of the experiment, $\mu \approx 12$. However, this value is inferred from an entirely classical observation of the total atom number, which does not require any quantum coherence to be generated during the interrogation time $T = 2.08\,$s. In order to confirm the quantum superposition principle at the same level of macroscopicity, one needs to detect the phase-stable interference fringes of a PS at 95\%  visibility. The difference between both scenarios becomes apparent if we compare them in terms of the amount of falsified MMM parameters, as given by the conditions $H(\sigma_q, \tau) < P$ versus $D(\sigma_q, \tau) < P$. Only the latter covers macroscopic $\tau$-values on the half-metre scale, $\hbar/\sigma_q \sim \Delta_x$, as it is shown in Fig. \ref{fig:exclusionplots}.

\subsection{Second-order correlations}\label{sec:dephasing2}

While single-particle observables can be used to demonstrate the phase-stable interference of a PS, they do not reveal possible quantum features of a DFS or PAPS as implied by the indistinguishability of the atoms. For this we must consider second-order correlations, i.e.~two-particle observables. 
Their modified time evolution can be solved explicitly if we neglect the free dispersion in the two arms, following the arguments of Sect.~\ref{sec:theory}. This amounts to integrating $\partial_t \la \op{O} \ra = \la {\cal L} (\op{O})\ra $ for any two-particle (or higher-order) observable $\op{O}$, see \ref{app:SecQuantCalc} for details.

For a PS of $N$ single-atom superpositions, we find the second-order correlations
\begin{eqnarray}
\fl \left<\op{c}_{a,{\rm out}}^{\dagger}\op{c}_{a,{\rm out}}^{\dagger}\op{c}_{a,{\rm out}}\op{c}_{a,{\rm out}}\right>_{\rm PS} &=& \frac{N(N-1)}{8} \left[3H\left(\sigma_q,\frac{\tau}{2}\right)+\cos(2\varphi)D\left(\sigma_q,\frac{\tau}{4}\right) \right. \nonumber \\ 
&& - 4\cos(\varphi)D(\sigma_q,\tau) \Big],
\label{Corr1PS}\\
\fl \left<\op{c}_{a,{\rm out}}^{\dagger}\op{c}_{b,{\rm out}}^{\dagger}\op{c}_{b,{\rm out}}\op{c}_{a,{\rm out}}\right>_{\rm PS} &=& \frac{N(N-1)}{8}
\left[H\left(\sigma_q,\frac{\tau}{2}\right)-\cos(2\varphi)D\left( \sigma_q,\frac{\tau}{4} \right) \right],
\label{Corr2PS}
\end{eqnarray}
with $\varphi = \phi - \alpha$. The probabilities of detecting two atoms in the same or in different output ports are thus affected by both MMM-induced depletion and dephasing. In Eq.~\eqref{Corr1PS}  the decay rate for the first order fringes, i.e.\ for the term proportional to $\cos(\varphi)$, is the same as in the first-order observable \eqref{eq:caca_PS}, whereas the depletion rate for the offset term doubles. The fact that the dephasing rate quadruples for the second-order fringes, which are proportional to $\cos(2\varphi)$, provides a way to gain macroscopicity from increased measurement resolution. 

The results for a PAPS are obtained by averaging the above terms over $\varphi$, they are thus insensitive to dephasing,
\begin{eqnarray}
\left<\op{c}_{a,{\rm out}}^{\dagger}\op{c}_{a,{\rm out}}^{\dagger}\op{c}_{a,{\rm out}}\op{c}_{a,{\rm out}}\right>_{\rm PAPS} =& \frac{3N(N-1)}{8} H\left(\sigma_q,\frac{\tau}{2}\right), \label{Corr1PAPS}\\
\left<\op{c}_{a,{\rm out}}^{\dagger}\op{c}_{b,{\rm out}}^{\dagger}\op{c}_{b,{\rm out}}\op{c}_{a,{\rm out}}\right>_{\rm PAPS} =& \frac{N(N-1)}{8} H\left(\sigma_q,\frac{\tau}{2}\right). \label{Corr2PAPS}
\end{eqnarray}
Comparing this to a balanced DFS \eqref{eq:DFS} with $N/2$ atoms in each arm, we find the same dependence on MMM-induced depletion,
\begin{eqnarray}
\left< \op{c}_{a,{\rm out}}^{\dagger} \op{c}_{a,{\rm out}}^{\dagger}\op{c}_{a,{\rm out}}\op{c}_{a,{\rm out}} \right>_{\rm DFS} =& \frac{N(3N - 2)}{8} H\left(\sigma_q,\frac{\tau}{2}\right) \label{Corr1DFS} \\
\left<\op{c}_{a,{\rm out}}^{\dagger}\op{c}_{b,{\rm out}}^{\dagger}\op{c}_{b,{\rm out}}\op{c}_{a,{\rm out}} \right>
_{\rm DFS}=& \frac{N(N-2)}{8} H\left(\sigma_q,\frac{\tau}{2}\right).
\label{Corr2DFS}
\end{eqnarray} 
The DFS and the PAPS differ by $N/8$ in the prefactor. Given the overall $N^2$ scaling, the relative difference is practically irrelevant for BEC experiments with $N\gg 1$ atoms. A clear distinction can only be made for small atom numbers, as illustrated in Fig.~\ref{secondordercorrelation}.

Our reasoning can be generalized in a straightforward manner to arbitrary correlations of order $K$ in the creation and annihilation operators associated to the two modes. Such $K$th-order correlations are given in terms of normally ordered products of $K$ creation and $K$ annihilation operators. (Other observables can be brought to normal order with help of the canonical commutation relations.) Taking the expectation value with respect to coherently split condensate states will result in a linear combination of phase-insensitive terms (or $K$-th moments of the populations in the two modes) and of coherence terms that depend on multiples of the relative phase, $k \varphi $ with $k \leq K$. Only the latter---which are absent in the case of DFS or PAPS---will be affected by MMM-induced dephasing and decay like $D(\sigma_q,\tau/k^2)$, at an enhanced rate of up to $k^2/\tau$ if $\sigma_q \Delta_x \gg \hbar$. The populations, on the other hand, will decay by MMM-induced depletion in each mode, as described by $H(\sigma_q,\tau/K)$. The linearly enhanced depletion rate saturates at $K/\tau$ for $\sigma_q w \gg \hbar$, with $w$ characterizing the three-dimensional Gaussian waist as in \eqref{eq:H_3D}. 

The quintessence of our findings is twofold. One the one hand, they show that a many-body enhanced macroscopicity or test of collapse and decoherence models (falling under the class of MMM) with BEC interference experiments requires the detection of genuine many-atom correlations. Otherwise, the BEC experiment is no more macroscopic than an equivalent single-atom interferometer at the same level of experimental uncertainty.
On the other hand, tests of the superposition principle by probing collapse models on the level of macroscopic arm separations and amplified decoherence rates can only be done with phase-coherent superposition states.

\subsection{Count statistics}\label{sec:measStatistics}

So far we have focused on the influence of MMM dephasing and depletion on expectation values of one- and two-particle observables. They are natural quantities to assess interference phenomena, but in the actual experiment they are derived from the raw sample of atom counts in the two output ports recorded over many runs. This data contains all the accessible information to distinguish different two-mode BEC states in terms of their quantum features and their sensitivity to MMM decoherence. 

In the following, we shall discuss the theoretical predictions for the atom count statistics, as described by the probabilities $P(n_a,N)$ of detecting $n_a$ atoms in output port $a$ and $n_b = N-n_a$ atoms in $b$, where $N$ is the initial (even) number of atoms in the condensate. In second quantization, the probabilities are given by $N$-particle expectation values at time $T$,
\begin{eqnarray}
P(n_a,N) = \frac{ \left<(\op{c}^{\dagger}_{a,{\rm out}})^{n_a}(\op{c}^{\dagger}_{b,{\rm out}})^{N-n_a}\left|{\rm vac}\right>\left<{\rm vac}\right|\op{c}_{a,{\rm out}}^{n_a}\op{c}_{b,{\rm out}}^{N-n_a}\right>}{{n_a!(N-n_a)!}}.
\label{eq:CountStatistics}
\end{eqnarray} 
As mentioned in the previous section and worked out in \ref{app:SecQuantCalc}, this $N$-th order expectation value comprises phase-independent terms and terms oscillating like multiples of the relative phase $\varphi$, depending on the properties of the condensate state. The former terms are only affected by MMM-induced depletion, whereas the latter by the dephasing effect. We see this explicitly after expanding \eqref{eq:CountStatistics} for the phase-coherent PS \eqref{eq:PS} in terms of the phase dependence, 
\begin{eqnarray}
\fl P_{\rm PS}(n_a,N) &=& \frac{1}{2^{2N}} \binom{N}{n_a}
\sum_{k,k'=0}^{n_a} (-)^{k-k'} \binom{n_a}{k}
\binom{n_a}{k'}
 \sum_{\ell,\ell'=0}^{N-n_a}
\binom{N-n_a}{\ell}
\binom{N-n_a}{\ell'} \nonumber\\
\fl && \times \left\{ \begin{array}{ll}
H\left(\sigma_q,\tfrac{\tau}{N}\right) & {\rm if}\,\, k+\ell=k'+\ell',\\
e^{i(k+\ell-k'-\ell')\varphi} D \left(\sigma_q,\tfrac{\tau}{|k+\ell-k'-\ell'|^2}\right) & {\rm else.}
\end{array} \right.
\label{messStatistikCSS}
\end{eqnarray} 

Like in Sect.~\ref{sec:dephasing2}, the dephasing rate grows as the square of the order of oscillation with $\varphi$. On the other hand, the depletion of all the phase-independent offset terms occurs at the same many-body amplified rate $N/\tau$, since the above probabilities refer to situations where all $N$ atoms remain in the condensate. 
The corresponding marginal probability is thus given by $\sum_{n_a} P_{\rm PS} (n_a,N) = H(\sigma_q, \tau/N)$.
\begin{figure}
  \centering
  \includegraphics[width=1\textwidth]{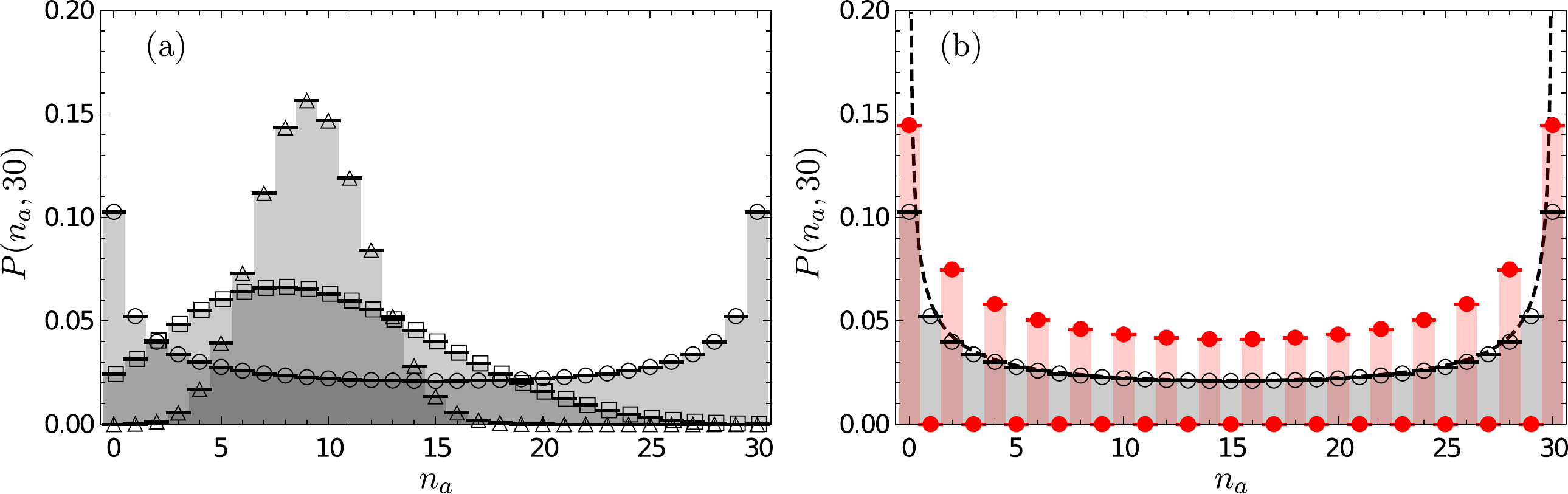}\\
  \caption{Atom count distribution $P(n_a,30)$ in output port $a$ for balanced two-mode BECs of 30 atoms subject to MMM-induced decoherence. a) We chose MMM parameters with a moderate dephasing strength \eqref{eq:D_MMM}, $\sigma_q \Delta_x / \hbar = 1$ and $T/\tau = 0.2$, while assuming negligible heating losses in the two arms, $H(\sigma_q,\tau/30) \approx 1$. The triangles and squares correspond to a PS with phase difference $\varphi=-3\pi/8$ in the absence and presence of dephasing, respectively. b) The open and the filled dots represent a PAPS and a DFS, both of  which are  unaffected by MMM. Notice the Hong-Ou-Mandel dips of the DFS, and the classical approximation (dashed line) for the PAPS.} 
 \label{messStatistik}
\end{figure} 

Figure~\ref{messStatistik} illustrates the effect of MMM dephasing on the count statistics for an exemplary PS containing $N=30$ atoms. We chose a parameter regime that results in moderate dephasing and negligible depletion losses. The peaked unmodified distribution (triangles) is smeared out (squares). At stronger dephasing, it will eventually approach the distribution of the associated PAPS (open dots). The statistics of both the PAPS and the corresponding DFS (filled dots) are not affected by the dephasing. 

This suggests that MMM dephasing can be probed directly by accumulating atom counts of a PS at a fixed phase and measuring the width of the count distribution $P(n_a=\langle\op{c}^{\dagger}_{a,{\rm out}}\op{c}_{a,{\rm out}}\rangle,N)$, thus evading the reconstruction of stable interference fringes from repeated measurements at varying phase shifts. For an initially pure PS of $N \gg 1$ atoms, the MMM-induced increase of the variance in $P(n_a,N)$ is given by 
\begin{align}
{\rm Var}(n_a) =&
\frac{N}{8}\Big[(N-3)H\left(\sigma_q, \frac{\tau}{2}\right)+(4+4N\cos\varphi\, D(\sigma_q,\tau))H\left(\sigma_q,\tau\right)\nonumber\\
&\vphantom{\frac{N}{8}}+(N-1)\cos(2\varphi)D\left(\sigma_q, \frac{\tau}{4}\right)-2N\cos^2\varphi\,D\left(\sigma_q, \frac{\tau}{2}\right)-4N\cos\varphi\,D(\sigma_q,\tau)\Big]\nonumber\\
\approx& \frac{N}{8}\left[(N-3)H\left(\sigma_q, \frac{\tau}{2}\right)+4H\left(\sigma_q,\tau\right)+(1-N)D\left(\sigma_q, \frac{\tau}{4}\right)\right]
\qquad\mbox{at $\varphi\approx\pm\pi/2$},
\label{eq:variance}
\end{align}
as follows by applying the results of Sects.~\ref{sec:dephasing1} and \ref{sec:dephasing2}. We see that the broadening depends on the same dephasing factor \eqref{eq:D_MMM} that reduces the interference fringe visibility. In the second line, we approximate the expression at the point of highest sensitivity to the MMM-induced broadening, $\varphi=\pm \pi/2$. There the variance can grow from $N/4$ to the maximum value $N(N+1)/8$ that corresponds to a PAPS. Measuring any value smaller than this maximum will place bounds on the dephasing strength. The black dotted curve in Fig.~\ref{fig:exclusionplots}, for example, marks the MMM parameters that would be ruled out by measuring a less than fourty-fold increase in the variance for $N=10^5$ atoms, given the experimental parameters used in Ref.~\cite{Kovachy2015}.  

The count statistics for the PAPS \eqref{eq:PAPS} follow by averaging the PS distribution  \eqref{messStatistikCSS} over the phase $\varphi$, 
\begin{eqnarray}
P_{\rm PAPS}(n_a,N)&=& \frac{H\left(\sigma_q,\tau/N \right)}{\pi n_a! (N-n_a)!}\Gamma \left( n_a + \tfrac{1}{2}\right) \Gamma \left( N-n_a + \tfrac{1}{2}\right) \nonumber \\
&\stackrel{N\gg 1}{\longrightarrow}& \frac{H\left(\sigma_q,\tau/N \right)}{\pi\sqrt{n_a(N-n_a)}}.
\end{eqnarray} 
We see that it is affected by the depletion effect only, uniformly  reducing  all probabilities at a given atom number $N$ by the constant prefactor $H$. 
For the many-atom limit $N\gg 1$, we have used Stirling's approximation of the gamma function. The depletion factor aside, the count distribution then approximates the  continuous classical expression $P_{\rm cl}(I_a|I) = 1/\pi\sqrt{I_a(I-I_a)}$; the latter describes the distribution of intensities $I_a$ expected for classical waves in one output port of a balanced beam splitter  when it receives two input beams of equal intensity $I/2$ but random phase difference \cite{Laloe2012}. In Fig.~\ref{messStatistik}, where  it is given by the dashed line, it approximates well the PAPS distribution for $N=30$. The classical expression is also used in the data analysis of Ref.~\cite{Kovachy2015}, see Eq.~(3) with $c=1$ there. A further smearing out of the distribution ($c<1$) might arise if atoms are lost or $N$ is not precisely known. 

For even $N$, the count distribution of the DFS can be written as 
\begin{eqnarray}
\fl P_{\rm DFS}(n_a,N) = H \left(\sigma_q,\frac{\tau}{N}\right) \times \left\{ \begin{array}{ll} \dfrac{(2j)! \Gamma(N/2-j+1/2)}{\sqrt{\pi} 2^{2j} (j!)^2 (N/2-j)!} & \text{if }n_a=2j \text{ even,} \\ 0 & \text{else,} \end{array} \right.
\label{DFSprobf}
\end{eqnarray}
which makes the Hong-Ou-Mandel effect explicit. The bosonic statistics of the condensate leads to constructive (destructive) interference for even (odd) atom counts in the detector. Once again, the distribution is  affected uniformly by depletion as it contains no phase coherence. In the limit of many atoms\footnote{Notice that Stirling's approximation is formally valid for $N-n_a, n_a \gg 1$ only. In practice, this leads to minor deviations at the margins.}, the Hong-Ou-Mandel dips of zero probability at odd atom numbers persist, and we obtain a modulation of the classical distribution,  $P_{\rm DFS} (2j,N\gg 1) \approx 2P_{\rm cl} (2j,N) H(\sigma_q, \tau/N)$, as seen in Fig.~\ref{messStatistik}. Distinguishing between a DFS and a PAPS in the experiment thus requires a detection efficiency and resolution on the single-atom level.

Finally, if phase coherence cannot be maintained and/or depletion losses were to be tested in an experiment, the data analysis should account for the fact that the initial or final atom number might not be precisely known. One can do this by averaging conditional count probabilities $P_{\rm PS}(n_a|N_d)$ that $N_d$ atoms are detected in total  to arrive at the unconditional distribution. The former is given by the corresponding \textit{undepleted} count probabilities $P_{\rm PS}(n_a|N_d)=[P_{\rm PS}(n_a,N_d)]_{H=1}$. 
For a product state, pure or phase-averaged, depletion can be modeled as a Bernoulli process characterized by the survival probability $H = H(\sigma_q,\tau)$,
\begin{equation}
P_{\rm PS}(n_a) = \sum_{N_d=0}^N \binom{N}{N_d}H^N_d (1-H)^{N-N_d} P_{\rm PS}(n_a|N_d).
\end{equation} 
An experimental scenario where the initial atom number $N$ is not precisely known can be described straightforwardly by a weighted average.

For a DFS, the unconditional count distribution follows from an average over two independent depletion processes removing atoms in each of the two independent condensates,
\begin{equation}
P_{\rm DFS}(n_a) = \sum_{N_a,N_b=0}^{N/2}P(n_a|N_a,N_b)\binom{N/2}{N_a}\binom{N/2}{N_b}H^{N_a+N_b} (1-H)^{N-N_a-N_b}. \label{eq:P_DFS_deplet}
\end{equation}
The undepleted distributions $P(n_a|N_a,N_b)$ for DFS populations $N_a,N_b \neq N/2$,  as well as their different Hong-Ou-Mandel signatures, are  examined in detail  in Ref.~\cite{Laloe2012}.

\begin{figure}
  \centering
  \includegraphics[width=1\textwidth]{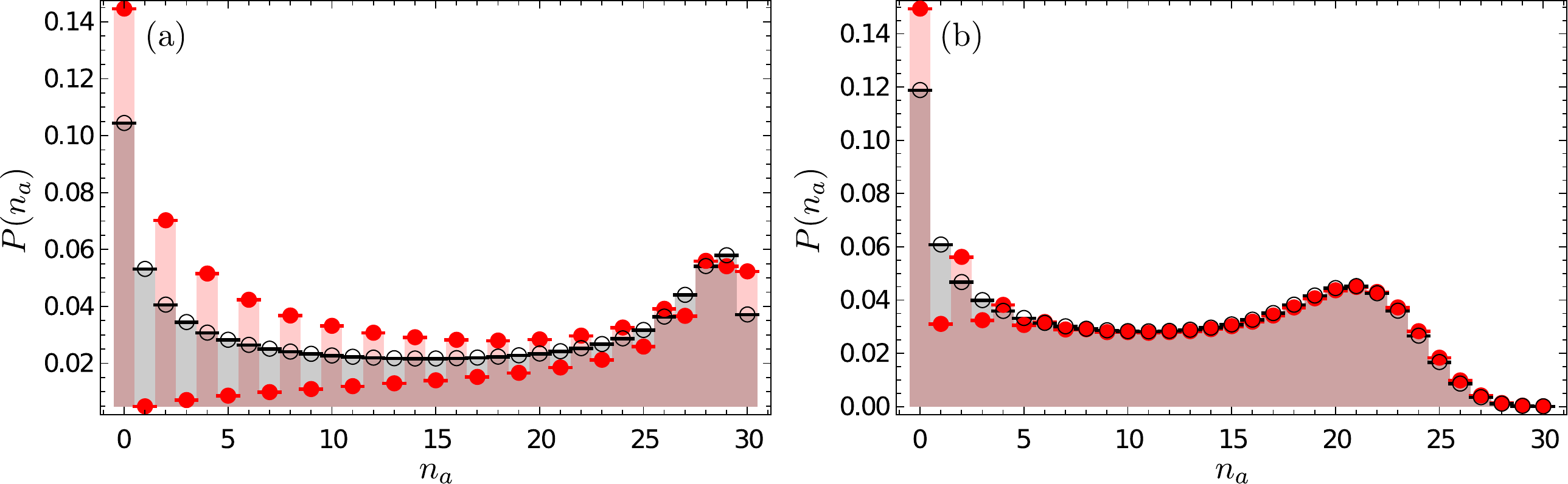}\\
     \caption{If the condensates loses a significant amount of particles due to heating, the HOM effect of the DFS (red full dots) gets partly destroyed. a) In  case that on average one particle is removed from the condensates  per run, the statistics for small $n_a$ are essentially unaffected, while the HOM dips vanish on the other side of the spectrum. b) Even if 25$\%$ of all particles are lost per shot, remnants of the HOM effect can verify a DFS. For comparison, the black open dots show the respective PAPS after the same amount of heating.}
\label{fig:DFSparticleloss}
\end{figure} 

The impact of particle loss on a DFS and a PAPS of 30 atoms is shown in Fig.~\ref{fig:DFSparticleloss}. Panel (a)  corresponds to an average loss of one atom, and  panel (b) to a loss of 25~\% of the atoms, $H=1-1/30$ and $H=0.75$, respectively. Compare this to the undepleted distributions in Fig.~\ref{messStatistik}(b). We observe that the count probabilities and in particular the HOM signature of the DFS prevail at the lower end of the count spectrum, whereas they are quickly diminished for higher count numbers $n_a$ rendering the DFS and the PAPS indistinguishable. Indeed, the relative frequency of detecting a maximum of $n_a = 30$ atoms in the experiment would be the most sensitive to depletion losses as it decays like $H^{30}$. 

Alternatively, by distinguishing between even and odd atom counts, the HOM signature of the DFS can be cast into a fairly simple quantum observable that comprises the full measurement record and achieves a comparable sensitivity. 
We find that the atom count number multiplied with its parity
serves this purpose well,
\begin{equation}
\langle n_a(-)^{n_a}\rangle 
= \sum_{n_a } n_a(-)^{n_a}\, P_{\rm DFS}(n_a)
\approx \frac{N}{2} H^{4N/3}(\sigma_q,\tau).
\label{eq:HOMobservable}
\end{equation}
This approximate dependence on the  depletion factor was inferred from a numerical evaluation of half-split DFS of $N \leq 100$ atoms, substituting \eqref{eq:P_DFS_deplet} for the count probabilities $P(n_a)$. Measuring a value greater than, say, 12 in an experiment with 30 atoms would exclude all $H(\sigma_q,\tau)<0.9944$, as represented by the black solid line in Fig.~\ref{fig:exclusionplots}. Note that such a sensitive measurement would require a stable preparation of a DFS with exactly 30 atoms in many runs of the experiment, as well as single-atom resolution in the detector.
This suggests that MMM-induced depletion can be effectively probed by interfering small (independent) condensates, at an exponentially growing sensitivity.

\section{Conclusions}\label{sec:concl}
We studied how a generic class of macrorealistic collapse theories and corresponding decoherence models affect superposition states of Bose-Einstein condensates in a two-mode Mach-Zehnder setting. The quantum signatures of these states, as typically detected by means of first- or second-order observables in the atom number, are then subject to dephasing and depletion, depending on the model parameters. By these means, the empirical macroscopicity of an interference experiment can be understood as the degree to which coherence-reducing processes are ruled out by the successful observation of quantum interference. This degree will thus depend on both the prepared many-body quantum state and on the measurement precision.

The interference of phase-coherent superposition states, i.e.~coherently split condensates, is in this sense always more macroscopic than the (second-order) interference of independent condensates with no fixed phase relation. While the former is vulnerable to both dephasing and depletion, the quantum signatures of the latter, which are a consequence of exchange symmetry, suffer only from depletion. However, depletion can also be probed in a purely classical scenario, e.g.~by monitoring atom losses in an incoherently split, or phase-averaged condensate. 
Sensing depletion and dephasing may even result in a comparable macroscopicity, if defined in terms of the greatest possible decoherence time ruled out by the experiment \cite{Nimmrichter2013}. The difference is that phase-coherent interferometry probes these decoherence times on the level of the possibly macroscopic arm separation, whereas the sensitivity of phase-incoherent experiments is limited to the size of the condensate in each arm.

By observing a stable interference contrast at low particle loss, an interference experiment with split atom condensates will never yield a higher macroscopicity than single-atom interference at a comparable number of repetitions with similar contrast and loss. We have shown that it is possible to exceed the single-atom sensitivity only by detecting higher-order phase fringes in the multi-particle correlation functions of the recombined condensate, which requires high-precision measurements in condensates of many atoms. Here future experiments may benefit from two-mode squeezing in particle number or phase if the goal is to improve the degree of macroscopicity. Number squeezing helps to achieve longer interference times by minimizing the phase dispersion effect related to atom interactions in dense condensate \cite{Jo2007,Schmiedmayer2013}, whereas phase squeezing can be used in dilute condensates to enhance the contrast of the higher-order fringe oscillations \cite{Ma2011,toth2012}. 

Using the empirical macroscopicity employed here as a yardstick, the latest single-atom and atomic condensate interferometers with their high phase sensitivity and long interrogation times are on par with state-of-the-art molecule interference experiments. This is remarkable since the latter realize NOON-type cat states of more than a hundred constituent atoms, whereas the former stay on the single-atom level. 
In this regard, one should keep in mind that the empirical measure of macroscopicity can be used to assess quantum states that may be viewed as  Schr\"odinger cats, but that it does not in itself serve to certify that a given state may be viewed as such.

\ack 
We thank Peter Asenbaum, Jason M.~Hogan, Mark A.~Kasevich, Tim Kovachy, and Dan M.~Stamper-Kurn for helpful discussions. 
S.N.~is supported by the National Research Foundation, Prime Minister's Office, Singapore, through the Competitive Research Programme (Award No. NRF-CRP12-2013-03) under the Research Centres of Excellence programme.

\appendix

\section{Mean effective arm separation for momentum-split condensates}\label{app:MomentumSplit}

In the main text, the two arms of the interferometer are represented by Gaussian wave packets with a static separation of $\Delta_x$. In practice, the incident condensate is initially split in momentum, diverges, gets reflected, and eventually recombines again at the output beam splitter. Here, we show the equivalence of both cases as far as MMM decoherence is concerned.

According to the single-particle description in Sect.~\ref{sec:macroWf}, the characteristic function of the split condensate can be expanded like \eqref{eq:chi2nd2mode}, given the Gaussian wavefunction $\psi_a (x)$ for mode $a$ and $\psi_b (x) = \exp (i\Delta_p x/\hbar) \psi_a (x)$. The characteristic function term associated to the coherence between both arms reads as
\begin{eqnarray}
M_{ab}(x,p)=\exp\left[-\frac{x^2}{8w_x^2}-\frac{(p-\Delta_p)^2w_x^2}{2\hbar^2} - \frac{i \Delta_p x}{2\hbar }\right],
\end{eqnarray}
replacing \eqref{eq:PabMab}. The population terms $P_{a,b} (x,p)$ need not be considered, since the MMM-induced heating will affect them in the same way as before. 

The time evolution of the condensates can be described by three consecutive steps.
First there is a free propagation over the time $T/2$. 
Then the mode $b$ gets reflected, which we can model as a momentum displacement by $-2\Delta_p$.
The third step is another propagation by $T/2$. 
Inserting the MMM-induced coherence decay, we arrive at the overall transformation
\begin{eqnarray}
\fl M_{ab}(x,p,T) &=& e^{i\Delta_p (x+pT/2m)/\hbar} M_{ab} \left(x,p \right) R\left(x+\frac{p-\Delta_p}{m} T,p - 2\Delta_p \right) R \left( x+\frac{pT}{2m},p \right), \nonumber \\
\fl R(x,p) &=& \exp \left[ \int_0^{T/2} \frac{\diff t}{\tau} e^{-\sigma_q^2 (x-pt/m)^2/2\hbar^2} -\frac{T}{2\tau} \right].
\end{eqnarray}
The momentum-split analogue of the coherence term \eqref{eq:erwM_MMM_1P} is then given by the overlap integral
\begin{eqnarray}
\fl \la \op{M}_{ab} (T) \ra \approx \la \psi_b | \rho_0 |\psi_a \ra \int \frac{\diff x\diff p}{2\pi \hbar} M_{ab}(x,p,T) M_{ab}^{*} (x,p) e^{-i\Delta_p (x+pT/2m)/\hbar},
\end{eqnarray}
assuming orthogonality of the two modes. Note that the integrand term $M_{ab}(x,p)$ is localized around $x=0$ and $p=\Delta_p$. If we once again assume a much larger typical arm separation than wave packet size, $\Delta_p T/2m \gg w_x$, and negligible dispersion, $t_d \ll T$, then we can approximate
\begin{eqnarray}
\la \op{M}_{ab} (T) \ra \approx \la \psi_b | \rho_0 |\psi_a \ra 
\exp \left\{ - \frac{T}{\tau} \left[ 1 - \frac{ \sqrt{2\pi} m\hbar}{\sigma_q \Delta_p T} {\rm erf} \left( \frac{\sigma_q \Delta_p T}{2\sqrt{2} m\hbar}\right) \right] \right\}.
\end{eqnarray}
Strictly speaking, the exponential MMM term replaces the dephasing function $D(\sigma_q,\tau)$ in \eqref{eq:erwM_MMM_1P} for momentum-split superpositions. However, the function also yields a maximum dephasing rate of $1/\tau$ in the limit of large $\sigma_q \gg m\hbar / \Delta_p T$, and a rate suppressed by the factor $ (\sigma_q \Delta_p T/m\hbar)^2/24 $ in the opposite limit. Hence we can describe the transition between both regimes approximately by the dephasing function \eqref{eq:D_MMM} with an effective static arm separation of $\Delta_x = \Delta_p T/ 2\sqrt{3} m$. Deviations from the dephasing function, limited to the regime where $ \sigma_q \Delta_x \sim \hbar $, are irrelevant when considering the range of experimentally excluded MMM parameters on the logarithmic scale in Fig. \ref{fig:exclusionplots}.

\section{Second quantization calculations}\label{app:SecQuantCalc}

Every $K$-th order correlation function $\left<\op{C}_K\right>$ is the expectation value of a normally ordered combination of the $K$ creation and $K$ annihilation operators in both output modes. Expanded in terms of field operators, $\op{c}_{j,\rm out} = \int\diff x\, \psi_{j,\rm out} (x) \hat{\psi}^{\dagger} (x)$, with the two wavefunctions $\psi_{a,\rm out} (x) = \left[ \psi_a (x) - \exp(i\alpha) \psi_b (x) \right]/\sqrt{2}$ and $\psi_{b,\rm out} (x) = \left[ \exp(-i\alpha) \psi_a (x) + \psi_b (x)\right]/\sqrt{2}$, we have
\begin{eqnarray}
\op{C}_K &=& : \prod_{k=1}^{K} \op{c}_{j_k,\rm out}^{\dagger} \op{c}_{j_{k+K},\rm out} :  = \int\diff x_1 \ldots \diff x_{2K} \, \op{F}_K \left(x_1, \ldots x_{2K} \right), \label{eq:CorrFkt0} 
\end{eqnarray}
with
\begin{eqnarray}
\op{F}_K \left(x_1, \ldots x_{2K} \right) &=& : \prod_{k=1}^K \psi_{j_k,\rm out} (x_k) \psi_{j_{k+K},\rm out}^{*} (x_{k+K}) \hat{\psi}^{\dagger} (x_k) \hat{\psi} (x_{k+K}): 
\end{eqnarray}
To obtain the MMM effect in the absence of dispersion, we apply the MMM generator (\ref{eq:MMM2nd}) to each normally ordered product $\op{F}_K$ of field operators. The resulting time evolution can be solved explicitly using the canonical commutation relations. We arrive at
\begin{eqnarray}
\fl \left<  \op{F}_K \left(x_1, \ldots x_{2K} \right) \right> = \left<  \op{F}_K \left(x_1, \ldots x_{2K} \right) \right>_0  \label{eq:CorrFktPsiSol} \\
\fl \quad \times \exp \left\{ \frac{T}{2\tau} \sum_{\mu,\nu =1}^K \left[ 2e^{-(x_{\mu}-x_{\nu+K})^2\sigma_q^2/2\hbar^2} - e^{-(x_{\mu}-x_{\nu})^2\sigma_q^2/2\hbar^2} - e^{-(x_{\mu+K}-x_{\nu+K})^2\sigma_q^2/2\hbar^2} \right] \right\}, \nonumber
\end{eqnarray}
where $\la \, \ra_0$ denotes the expectation value predicted in the absence of MMM.

Computing the expectation value of $\op{F}_K$ for a DFS with $N_a+N_b$ atoms in the two modes, one finds that only those combinations of wavefunctions survive where the coordinates $x_{\mu}$ are all localized in one of the arms. In the dephasing regime $\sigma_q w_x \ll \hbar$ we expand the Gaussians in the exponent of \eqref{eq:CorrFktPsiSol} to second order. Together with the Gaussian wavefunctions, this can then be inserted back into \eqref{eq:CorrFkt0}. We express the resulting multidimensional Gaussian integral in terms of the $2K\times2K$ matrix
\begin{eqnarray}
\mathbb{M}_K=\frac{\sigma_q^2 T}{\hbar^2\tau}
\begin{pmatrix}
1+\frac{\hbar^2\tau}{w_x^2\sigma_q^2T}	& 1	& \dots	 & -1 & -1     \\
1	& 1+\frac{\hbar^2\tau}{w_x^2\sigma_q^2T} & \dots  & -1 & -1	  \\
\vdots & \vdots & \ddots & \vdots & \vdots\\
-1  & -1 & \dots	& 1+\frac{\hbar^2\tau}{w_x^2\sigma_q^2T} & 1 \\
-1 	& -1 & \dots & 1	 & 1+\frac{\hbar^2\tau}{w_x^2\sigma_q^2T}
\end{pmatrix}.
\end{eqnarray}
Integration yields the depletion term 
\begin{eqnarray}
\frac{\la \op{C}_K \ra_{\rm DFS} }{\la \op{C}_K \ra_{0,\rm DFS}} &=& \int \frac{\diff x_1 \dots \diff x_{2K}}{(2\pi w_x^2)^K} 
\exp\left[-\frac{1}{2}\vect{x}\cdot\left(\mathbb{M}_K\vect{x}\right)\right]
=\frac{1}{w_x^{2K}\left| \mathbb{M}_K \right|^{1/2}}  \nonumber \\
&=&\left(1+ 2 K  \frac{w_x^2\sigma_q^2 T}{\hbar^2\tau}\right)^{-1/2} \approx H \left( \sigma_q, \frac{\tau}{K} \right).
\end{eqnarray}
Expressing the result in terms of the function $H$ accounts also for the opposite regime of strong depletion, $\sigma_qw_x\gg\hbar$, where the $K$-th order decay rate saturates at $K/\tau$. 

In particular, the second-order correlations of a DFS read as
\begin{eqnarray}
\fl \left< \op{c}_{a,{\rm out}}^{\dagger} \op{c}_{a,{\rm out}}^{\dagger}\op{c}_{a,{\rm out}}\op{c}_{a,{\rm out}} \right>_{\rm DFS} &=& H\left(\sigma_q,\frac{\tau}{2}\right) \frac{N_a(N_a-1)+4N_a N_b+N_b(N_b-1)}{4}, \label{Corr1DFSgen}\\
\fl \left<\op{c}_{a,{\rm out}}^{\dagger}\op{c}_{b,{\rm out}}^{\dagger}\op{c}_{b,{\rm out}}\op{c}_{a,{\rm out}} \right>
_{\rm DFS} &=& H\left(\sigma_q,\frac{\tau}{2}\right) \frac{N_a(N_a-1)+N_b(N_b-1)}{4},
\label{Corr2DFSgen}
\end{eqnarray}
see \eqref{Corr1DFS} and \eqref{Corr2DFS} for the balanced case $N_a=N_b=N/2$.

A similar calculation can be done for a PS. In this case, $\la \op{F}_K \ra $ will consist of products of $K$ arbitrary pairs of mutually conjugate wavefunctions. Repeating all previous approximations, one finds that if a term contains $k\leq K$ mixed pairs $\psi_a \psi_b^{*}$ or $\psi_a^{*} \psi_b$, it will dephase like $D(\sigma_q,\tau/k^2)$, i.e.~at a quadratically enhanced rate. The remaining terms with no mixed pairs are subject to the linearly enhanced depletion $H(\sigma_q,\tau/K)$ as before.

\section*{References}

\providecommand{\newblock}{}

\end{document}